\documentclass[aps,prc,preprint,groupedaddress,showpacs]{revtex4-1}

\usepackage{graphicx}
\usepackage{amsmath,bm,ascmac,amssymb,calc}
\usepackage{CJK}

\begin{document}
\begin{CJK*}{UTF8}{}
\CJKfamily{min}


\title{Effects of realistic tensor force on nuclear quadrupole deformation
around the shore of the island of inversion}


\author{Y. Suzuki\,(鈴木優香)}
\author{H. Nakada\,(中田仁)}
\email[E-mail:\,\,]{nakada@faculty.chiba-u.jp}
\author{S. Miyahara\,(宮原聡)}

\affiliation{Department of Physics, Graduate School of Science,
 Chiba University,\\
Yayoi-cho 1-33, Inage, Chiba 263-8522, Japan}


\date{\today}

\begin{abstract}
The M3Y-type semi-realistic interaction is applied to deformed nuclei
for the first time.
The constrained Hartree-Fock calculations assuming axial symmetry
are implemented for the $N=20$ isotones $^{30}$Ne, $^{32}$Mg, $^{34}$Si
and the $N=28$ isotones $^{40}$Mg, $^{42}$Si, $^{44}$S with the M3Y-P6 interaction.
The results match the experimental data well.
Effects of the realistic tensor force on the nuclear quadrupole deformation
are investigated
in relation to the loss of the $N=20$ and $28$ magic numbers.
The tensor force is confirmed to favor the deformation for the $N=28$ nuclei
owing to the closure of the $jj$-shell (\textit{i.e.}, $n0f_{7/2}$),
while favoring the sphericity for the $N=20$ nuclei
owing to the $\ell s$-closure of $N=20$.
\end{abstract}


\maketitle
\end{CJK*}



\section{Introduction\label{sec:intro}}

The tensor force is an important ingredient of the interaction among nucleons.
This force is naturally derived in the meson exchange picture
for the nucleonic interaction,
and its presence is verified by the non-vanishing quadrupole moment
of the deuteron.
However, although some of its effects may be incorporated
in the effective interaction,
explicit inclusion of the tensor force had been avoided
in the self-consistent nuclear structure calculations.
Only relatively recently the tensor force has been recognized
to play a significant role in the $Z$- or $N$-dependence
of the shell structure~\cite{ref:Vtn,ref:SC14}.
The tensor force may be responsible for the appearance and disappearance
of magic numbers~\cite{ref:Vtn,ref:NS14},
which have been revealed in experiments
utilizing the radioactive beams~\cite{ref:SP08}.

While the effects of the tensor force on the nuclear mean field (MF)
have been argued for spherical nuclei in a number of recent studies,
investigations of its effect on nuclear deformation have remained limited.
Bender \textit{et al.}~\cite{ref:Sky-TNS} applied
the Skyrme energy density functional (EDF)
that contains the terms connected to the tensor force
to the quadrupole deformation,
and analyzed the influence of these terms in detail.
Although this treatment had already clarified some important aspects
of the tensor force related to the nuclear deformation,
the real effects of the tensor force remain elusive,
primarily because the relevant EDF terms are fully parameterized,
with signs and strengths not well determined.
It should be noted that the Skyrme EDF is constrained to the quasi-local form,
and the relevant terms in it are not well distinguished
from the contribution of the central force.

One of the authors (H.N.) has developed
semi-realistic nucleonic interactions~\cite{ref:Nak03,ref:Nak13,ref:NI15},
which are based on the $G$-matrix approach~\cite{ref:M3Y-P}
but are modified from the phenomenological viewpoints.
It has been found that the magic numbers are well described
by the M3Y-P6 interaction
throughout a wide range of the nuclear chart
including unstable nuclei~\cite{ref:NS14}.
Owing to the realistic tensor force in it,
the semi-realistic interaction will be useful
for investigating the tensor-force effects on nuclear deformation.
Conventional MF calculations, not featuring tensor force,
have been successful in describing many significant aspects
of quadrupole deformation at and around the $\beta$ stability~\cite{ref:SCMF}.
However, unable to satisfactorily reproduce the $Z$- and $N$-dependence
of the shell structure,
the reliability of these calculations for nuclei far off the $\beta$-stability,
\textit{e.g.} those in the so-called island of inversion~\cite{ref:WBB90},
might be questioned.
In practice, it has not been easy to capture deformation
for $^{30}$Ne and $^{32}$Mg~\cite{ref:Ter97,ref:Peru00,ref:RG02},
although correlations beyond the MF framework may contribute~\cite{ref:KH02}
and interpretation other than deformation
has not fully been excluded~\cite{ref:YG04}.
In the present work,
we apply the semi-realistic M3Y-P6 interaction
to the Hartree-Fock (HF) calculations assuming the axial symmetry,
for the nuclei located near the shore (\textit{i.e.} the borderline)
of the island of inversion;
three $N=20$ isotones ($^{30}$Ne, $^{32}$Mg and $^{34}$Si)
and three $N=28$ isotones ($^{40}$Mg, $^{42}$Si and $^{44}$S).
This is the first application of the M3Y-type semi-realistic interaction
to deformed nuclei.
We then investigate the tensor-force effects on quadrupole deformation
in these nuclei.
Although we leave to future work
inclusion of the pair and rotational correlations
via the Hartree-Fock-Bogolyubov (HFB) calculations
and the angular-momentum projection (AMP),
we note that these correlations are basically irrelevant to the tensor force.
Being free of complications which may be brought in by these correlations,
our HF calculations are useful for studying the tensor-force effects
on nuclear deformation.

\section{Effective Hamiltonian and numerical method\label{sec:Hamil}}

We have implemented self-consistent HF calculations
with the non-relativistic effective Hamiltonian $H=H_N+V_C-H_\mathrm{c.m.}$,
which is rotationally and translationally symmetric
except the density-dependence.
The nuclear part $H_N$,
\begin{equation}
H_N = K + V_N\,;\quad K = \sum_i \frac{\mathbf{p}_i^2}{2M}\,,\quad
V_N = \sum_{i<j} v_{ij}\,,
\label{eq:H_N}\end{equation}
where $i$ and $j$ represent the indices of individual nucleons,
is taken to be isoscalar.
The term $V_C$ represents the Coulomb interaction between protons.
Consistent with $K$, the center-of-mass (c.m.) part
is $H_\mathrm{c.m.}=\mathbf{P}^2/2AM$,
with the total momentum $\mathbf{P}=\sum_i \mathbf{p}_i$
and the mass number $A\,(=Z+N)$.
The effective nucleonic interaction is
\begin{equation} v_{ij} = v_{ij}^{(\mathrm{C})}
 + v_{ij}^{(\mathrm{LS})} + v_{ij}^{(\mathrm{TN})}
 + v_{ij}^{(\mathrm{C}\rho)}\,,
\label{eq:effint}\end{equation}
where $v^{(\mathrm{C})}$, $v^{(\mathrm{LS})}$ and $v^{(\mathrm{TN})}$
are two-nucleon interaction in the central, LS and tensor channels,
all of which have the finite-range Yukawa form in the M3Y-type interactions.
$v^{(\mathrm{C}\rho)}$ is the density-dependent central force,
which is significant to give appropriate saturation properties~\cite{ref:Nak03}.
We mainly use the M3Y-P6 parameter-set~\cite{ref:Nak13} in this work.
Although a density-dependent LS channel was introduced
in Refs.~\cite{ref:NI15,ref:Nak15},
we here use $v_{ij}^{(\mathrm{LS})}$ which is density-independent.
In the M3Y-P$n$ interactions with $n\geq 5$~\cite{ref:Nak13,ref:Nak10},
$v^{(\mathrm{TN})}$ is kept unchanged
from the M3Y-Paris interaction~\cite{ref:M3Y-P}
that was obtained from the $G$-matrix.
A detailed description of M3Y-P6 is provided in Ref.~\cite{ref:Nak13}.
Note that both the one- and two-body terms in $H_\mathrm{c.m.}$
are subtracted before iteration,
and the Coulomb exchange term is handled precisely.

In numerical calculations,
we have employed the Gaussian expansion method~\cite{ref:GEM},
which has been extended to the axially deformed MF calculations
in Ref.~\cite{ref:Nak08}.
Single-particle (s.p.) functions under axially deformed one-body fields
are expressed by superposition of spherical Gaussian basis-functions.
Whereas the spherical bases are impractical for describing
very large deformations such as the superdeformation,
they are suitable for arguing structure of deformed nuclei
in terms of spherical orbits.
Numerical errors owing to the truncation with respect to $\ell$
(orbital angular momentum) are
estimated as $\lesssim A^{1/3}\,\mathrm{MeV}$~\cite{ref:Nak08}
for normally deformed states studied here.

To obtain the deformed energy minima,
axially deformed configurations should be given as initial conditions.
In the present calculations,
we have set initial configurations by the deformed Woods-Saxon (WS) potential,
assuming a certain value for the deformation parameter $\beta$,
and then have solved the HF equation iteratively until ensuring convergence.
Several minima have been searched for each nucleus by varying $\beta$
of the initial WS potential, from $-0.9$ to $0.9$ in steps of $0.1$.
Notice that the minimum of the total energy does not necessarily imply
that the s.p. levels should be filled in the order of s.p. energies.
When inverted or closely located s.p. levels are present
around the Fermi energy,
we have conducted additional calculations
by purposely filling the other configuration,
and have compared the total energies.
Moreover, we have implemented the constrained HF (CHF) calculations
by adding the quadratic constraining term to the Hamiltonian
as $\tilde{H}=H+c(\hat{q}_0-d)^2$,
where $\hat{q}_0$ is the mass quadrupole moment operator defined by
$\hat{q}_0=\sqrt{(16\pi/5)}\sum_i r_i^2\,Y^{(2)}_0(\hat{\mathbf{r}}_i)$.
The parameter $c$ has been fixed to $0.01\,\mathrm{MeV}\,\mathrm{fm}^{-2}$
in all CHF calculations in the present work.
Starting from each energy minimum of $H$,
we have gradually varied the value of $d$.
The variational calculations with respect to $\langle\Phi|\tilde{H}|\Phi\rangle$
have been implemented for fixed $c$ and $d$,
by which a Slater determinant $|\Phi\rangle=|\Phi(q_0)\rangle$
has been determined.
The gradient method has been used in these CHF calculations,
starting from the solution at adjacent $d$.
Energy curves $E(q_0)$ are plotted
(\textit{i.e.}, $E=\langle\Phi|H|\Phi\rangle$
as a function of the mass quadrupole moment $q_0$),
for $|\Phi\rangle\,(=|\Phi(q_0)\rangle)$
that minimizes $\langle\Phi|\tilde{H}|\Phi\rangle$.
We here obtain $q_0$ from
\begin{equation}
 q_0 = \sqrt{\frac{16\pi}{5}}\langle\Phi\big|
 \sum_i r_i^{\prime 2}\,Y^{(2)}_0(\hat{\mathbf{r}'}_i)\big|\Phi\rangle
 = \langle\Phi|\hat{q}_0|\Phi\rangle
 - \sqrt{\frac{16\pi}{5}}\langle\Phi|R^2\,Y^{(2)}_0(\hat{\mathbf{R}})
 |\Phi\rangle\,,
\end{equation}
where $\mathbf{r}'_i=\mathbf{r}_i-\mathbf{R}$
with $\mathbf{R}=(1/A)\sum_i\mathbf{r}_i$.
Influence of the c.m. motion on the quadrupole moment is removed in $q_0$,
while it is ignored in $\tilde{H}$ for the sake of simplicity.

We have also computed the expectation value $E^{(\mathrm{TN})}
= \langle\Phi|\sum_{i<j}v_{ij}^{(\mathrm{TN})}|\Phi\rangle$,
and plot $E(q_0)-E^{(\mathrm{TN})}(q_0)$ to reveal tensor-force effects.
One might think that the tensor-force effects are obtained
via the minimization of $\langle H-\sum_{i<j}v_{ij}^{(\mathrm{TN})}\rangle$,
rather than the expectation value with $|\Phi\rangle$
that minimizes $\langle H\rangle$ (or $\langle\tilde{H}\rangle$).
However, since several parameters in M3Y-P6 have been adjusted
under the presence of $v^{(\mathrm{TN})}$,
it is not clear what the state stands for
that is obtained from the minimization
of $\langle H-\sum_{i<j}v_{ij}^{(\mathrm{TN})}\rangle$.
In most of the present cases, the effects of $v^{(\mathrm{TN})}$ are perturbative
and the minimization of $\langle H-\sum_{i<j}v_{ij}^{(\mathrm{TN})}\rangle$
almost coincides with $E(q_0)-E^{(\mathrm{TN})}(q_0)$,
as will be shown later.
The axial CHF calculations have been also performed
by adopting the Gogny-D1M interaction~\cite{ref:D1M}
for $V_N$ in Eq.~(\ref{eq:H_N}),
which is one of the most successful effective interactions
without including the tensor force.
Although other channels may also contribute,
a comparison with the D1M results suggests
difference between interactions fitted with and without the tensor force.

\section{Energy curves\label{sec:energy}}

In this section we present $E(q_0)$ for the $N=20$ and $28$ nuclei
and discuss loss or persistence of the $N=20$ and $28$ magic numbers in them.
It should be commented, however,
that the definition of `magicity' is not necessarily obvious.
Theoretically, magicity is related to a large shell gap
and resultant strong quenching of many-body correlations.
Therefore one of the reasonable theoretical definitions
is given by the stability of the spherical HF configuration
with respect to these correlations.
Although it is not easy to compare the shell gap and the correlations
quantitatively,
pairing was used as a measure of the correlations in Ref.~\cite{ref:NS14},
by performing the spherical HFB calculations.
The quadrupole deformation is another important correlation
that may violate magicity.
By implementing the axial HF calculations,
persistence and loss of the $N=20$ and $28$ magicities
against the quadrupole deformation are investigated,
and the role of the tensor force in determining them is discussed.
The neutron configuration in the ground state is
$(0d_{5/2})^6(1s_{1/2})^2(0d_{3/2})^4$
[$(0d_{5/2})^6(1s_{1/2})^2(0d_{3/2})^4(0f_{7/2})^8$]
if the $N=20$ ($N=28$) magicity is maintained.
This configuration favors a spherical shape.
As a deviation from this configuration leads to a certain deformation
within the HF framework,
the extent of the deformation is expected to be a measure of the magicity.

\subsection{$N=20$ nuclei: $^{30}$Ne, $^{32}$Mg and $^{34}$Si}

Before proceeding to discuss the nuclei around the island of inversion,
we show, in Table~\ref{tab:min-N20a},
$q_0$ and $E$ for the lowest and second lowest energy minima
of the $\beta$-stable $N=20$ nuclei $^{36}$S, $^{38}$Ar and $^{40}$Ca.
The deformation parameter $\beta$ is also presented,
which is extracted from $q_0$ via the following relation~\cite{ref:BM2},
\begin{equation}
q_0 = \frac{3\beta}{\sqrt{5\pi}} AR^2\,;\quad R=1.12A^{1/3}\,\mathrm{fm}\,.
\end{equation}
Another deformation parameter $\delta$ is sometimes used~\cite{ref:BM2}.
Note that, for instance,
$\beta\approx 0.5$ corresponds to $\delta\approx 0.35$ in this mass region.
The results of the spherical HFB calculations with M3Y-P6~\cite{ref:NS14}
and the measured ground-state energies~\cite{ref:AME12} are given for reference.
It is confirmed that the spherical minimum,
at which the $N=20$ magicity is preserved,
is well developed in $^{36}$S and $^{40}$Ca.
In $^{38}$Ar there are two minima with very small deformation.
This deformation occurs because protons occupy $0d_{3/2}$ only partially,
by which $q_0=0$ is prohibited within axial HF.
These minima should merge into a spherical state
once the pair correlation is set on as in the HFB.

\begin{table}
\caption{Intrinsic mass quadrupole moment $q_0$ ($\mathrm{fm^2}$),
deformation parameter $\beta$ and energy $E$ ($\mathrm{MeV}$)
at the lowest and second lowest minima
obtained in the axial HF calculations,
for the $\beta$-stable $N=20$ nuclei.
Results of the M3Y-P6 interaction are compared
with those of the D1M interaction.
For reference,
the energy obtained from the spherical HFB calculation with M3Y-P6 (5th column)
and the measured energy of the ground state (right most column)~\cite{ref:AME12}
are also presented.
\label{tab:min-N20a}}
\begin{ruledtabular}
\begin{tabular}{crrrrrrrr}
nuclide & \multicolumn{4}{c}{M3Y-P6} & \multicolumn{3}{c}{D1M} & Exp.~~\\
& $q_0$~ & $\beta$~~ & $E$~~~ & $E_\mathrm{HFB}$~ & $q_0$~ & $\beta$~~ & $E$~~~
 & $E_{g.s.}$~~\\
\hline
$^{36}$S & $0$ & $0.00$ & $-300.3$ & $-301.1$ & $0$ & $0.00$ & $-305.1$
 & $-308.7$ \\
 & $-139$ & $-0.37$ & $-294.8$ && $141$ & $0.38$ & $-299.3$ &\\
$^{38}$Ar & $-31$ & $-0.08$ & $-319.0$ & $-320.3$ & $-32$ & $-0.08$ & $-324.5$
 & $-327.3$ \\
 & $23$ & $0.06$ & $-318.9$ && $21$ & $0.05$ & $-324.4$ &\\
$^{40}$Ca & $0$ & $0.00$ & $-335.9$ & $-336.0$ & $0$ & $0.00$ & $-342.2$
 & $-342.1$ \\
 & $-177$ & $-0.40$ & $-327.5$ && $-170$ & $-0.38$ & $-330.7$ &\\
\end{tabular}
\end{ruledtabular}
\end{table}

\begin{figure}
\includegraphics[scale=1.0]{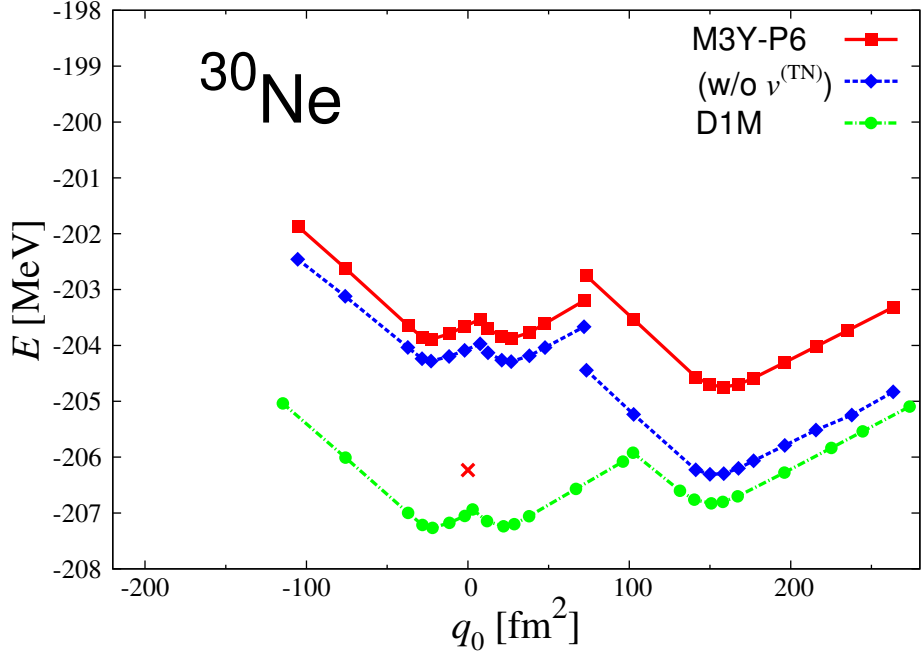}\vspace*{1cm}
\caption{(Color online) CHF results
of $E(q_0)$ (red squares) and $E(q_0)-E^{(\mathrm{TN})}(q_0)$ (blue diamonds)
for $^{30}$Ne, which are obtained with M3Y-P6.
For comparison,
the energy obtained from the spherical HFB calculation (red cross)
and $E(q_0)$ with D1M (green circles) are also plotted.
Lines are drawn to guide the eye.
\label{fig:Ne30_E-q0}}
\end{figure}

\begin{figure}
\includegraphics[scale=1.0]{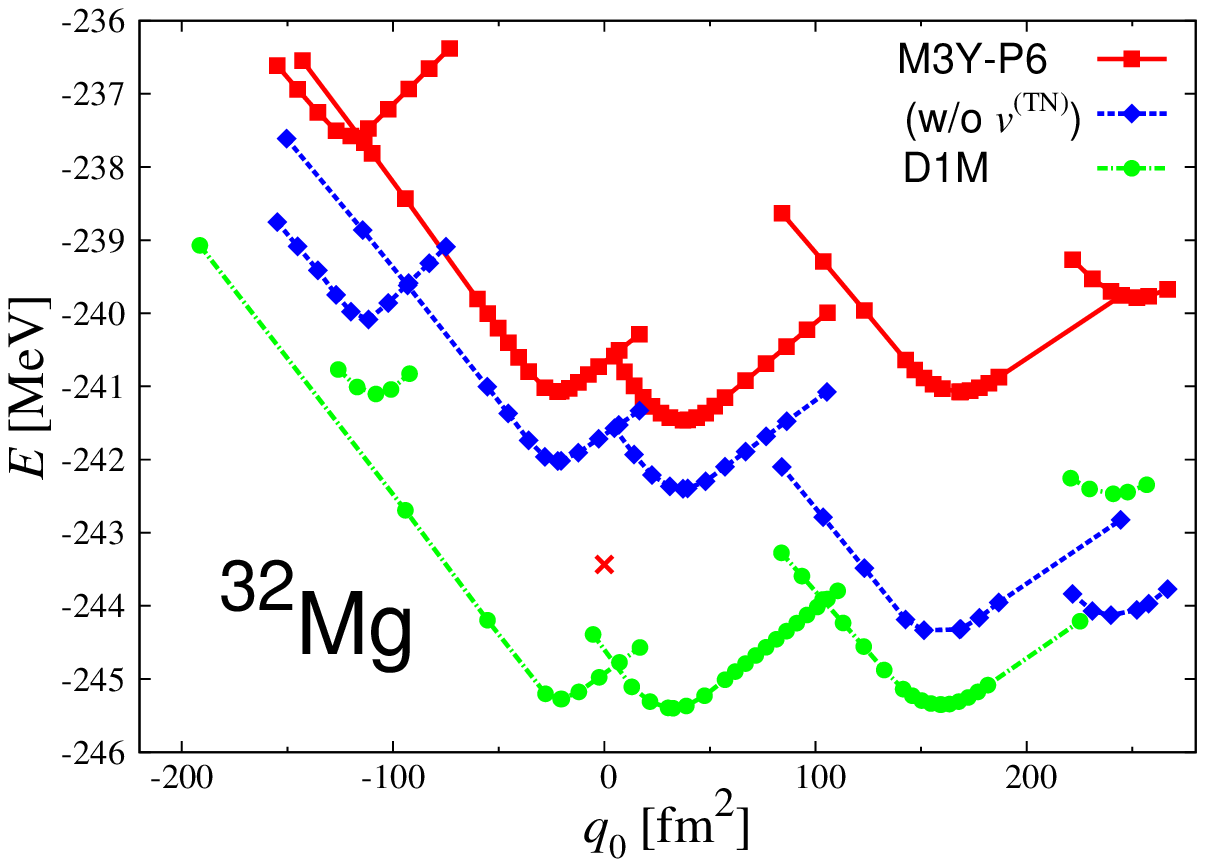}\vspace*{1cm}
\caption{(Color online) CHF results for $^{32}$Mg.
See Fig.~\protect\ref{fig:Ne30_E-q0} for conventions.
\label{fig:Mg32_E-q0}}
\end{figure}

\begin{figure}
\includegraphics[scale=1.0]{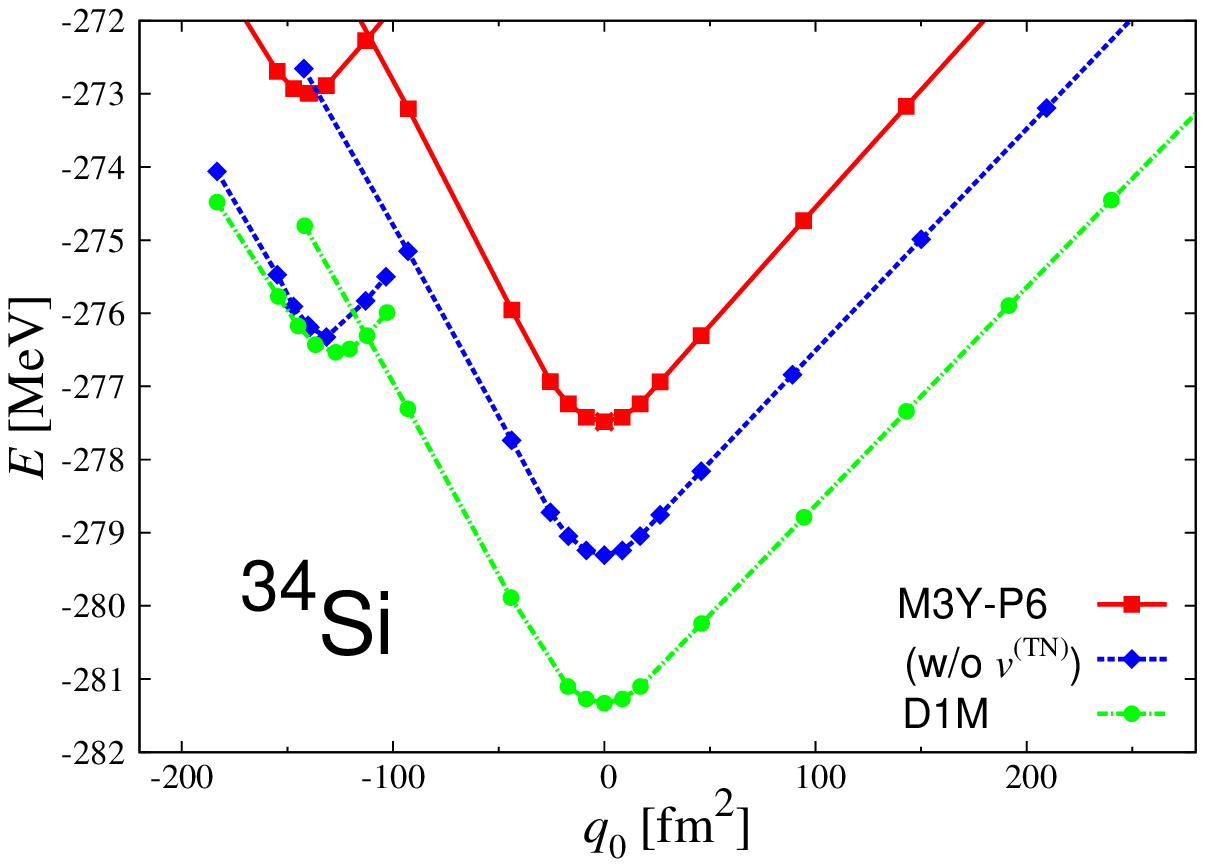}\vspace*{1cm}
\caption{(Color online) CHF results for $^{34}$Si.
See Fig.~\protect\ref{fig:Ne30_E-q0} for conventions.
Notice that the spherical HFB energy coincides with $E(q_0=0)$.
\label{fig:Si34_E-q0}}
\end{figure}

The energy curves of $^{30}$Ne, $^{32}$Mg and $^{34}$Si
obtained from the CHF calculations are depicted
in Figs.~\ref{fig:Ne30_E-q0}, \ref{fig:Mg32_E-q0} and \ref{fig:Si34_E-q0},
respectively.
In addition to $E(q_0)$ and $E(q_0)-E^{(\mathrm{TN})}(q_0)$ with M3Y-P6,
$E(q_0)$ with D1M and the spherical HFB energy (with M3Y-P6)
are shown for comparison.

For each of the three nuclei,
the minima are obtained at similar $q_0$, irrespective of interactions.
This suggests that, at the HF level, the $q_0$ value yielding a local minimum
is determined primarily by the configuration,
almost independent of the effective interaction.
On the other hand, the energies at these local minima and their difference
significantly depend on the interactions.
As a result, the $q_0$ value at the absolute minimum
also depends on the interaction.
It is noted that, in most cases, different configurations
do not mix one another because of the axial symmetry of the HF field,
as will be discussed in more detail in Sec.~\ref{sec:s.p.level}.

Let us first consider the case of $^{34}$Si,
presented in Fig.~\ref{fig:Si34_E-q0}.
In both the results of D1M and M3Y-P6,
a well-developed spherical minimum is found at $q_0=0$.
Together with the suppression of the pair correlation~\cite{ref:NS14},
this suggests that $^{34}$Si may behave as a doubly-magic nucleus.
It is recalled that $^{34}$Si has been considered
as a candidate nucleus for the proton bubble structure~\cite{ref:NSM13},
in which non-occupancy of the $p1s_{1/2}$ orbit causes
depletion of the proton density at the nuclear center.
If the nucleus were deformed,
there should be admixture of the orbital angular momenta
in the s.p. functions,
which may break the bubble structure
because it indicates partial occupation of $p1s_{1/2}$.
Since the deformation is unlikely,
the present calculation further supports the bubble structure in this nucleus.

Comparison between $E$ and $E-E^{(\mathrm{TN})}$ reveals
contribution of the tensor force.
The tensor force is repulsive for all $q_0$.
This feature is generic as long as we focus on the energy curves,
as shall be argued later.
A local minimum with oblate deformation is observed,
with significantly higher energy than the spherical minimum.
The repulsive effect of the tensor force is stronger at this minimum
than at the spherical minimum.
Thereby the sphericity in $^{34}$Si is enhanced by the tensor force.
It is interesting to observe that the energy difference
between the spherical and oblate minima obtained with D1M
is close to that with M3Y-P6 which includes $v^{(\mathrm{TN})}$.
In this respect the tensor-force effects are effectively incorporated
in the parameters of D1M.

Five local minima are observed in $^{32}$Mg.
Two of them are located around $q_0=0$.
This nucleus cannot be completely spherical in the axial HF calculations,
since protons occupy $0d_{5/2}$ only partially.
In these two minima the $N=20$ magicity is almost maintained,
and these minima are expected to merge and form a spherical solution
after the pair correlation is set on.
In contrast, the $N=20$ magicity is broken
at the minimum with $q_0\approx 170\,\mathrm{fm}^2$,
as confirmed by occupation of the $\Omega^\pi=1/2^-$ level
originating from the $n0f_{7/2}$ orbit
(see Fig.~\ref{fig:Mg32_spe} and related arguments).
Both in the D1M and M3Y-P6 results,
this deformed minimum is competing in energy with the nearly spherical minima.
If we consider correlations beyond the HF approximation,
the spherical configuration should gain energy from the pair correlations,
while the energy of the deformed one should be lowered
owing to the rotational correlations that may be evaluated by the AMP.
It should not be claimed which minimum yields the ground state,
until computations taking account of both the pairing and the rotation
are implemented.
We just point out that the close energies may indicate
the possibility of the shape coexistence at a low energy,
which has been suggested by experiments~\cite{ref:Mg32-0+}.

In Ref.~\cite{ref:NS14},
it has been shown that the $N=20$ magicity is not destroyed in $^{32}$Mg
within the spherical HFB calculation with M3Y-P6.
The present result illustrates that,
although the spherical HFB is useful for searching candidates of magic numbers,
the magicity could be lost via the quadrupole deformation.

In $^{30}$Ne, three local minima are observed.
Whereas the $N=20$ magicity appears to hold for two of them,
it is broken for the prolate minimum with $q_0\approx 160\,\mathrm{fm}^2$.
Although the prolate minimum is slightly higher
than the minima around the sphericity in the D1M result,
it is lower in the M3Y-P6 result.
It is of interest how and to what extent these results are affected
when the pair and rotational correlations are taken into account.
In the M3Y-P6 result, the energy of the deformed minimum goes up
and becomes closer to those of the nearly spherical minima
by the tensor force,
but is still lower than those of the minima near the sphericity.

The values of $q_0$, $\beta$ and $E$ at the lowest and the second lowest minima
are listed in Table~\ref{tab:min-N20} for individual nucleus.
The spherical HFB energies with M3Y-P6~\cite{ref:NS14}
and the measured ground-state energies~\cite{ref:AME12} are given for reference.
Because various correlations, including the pair and the rotational ones,
lower the energies to a certain extent,
it is reasonable that the HF energies are higher
than the measured ground-state energies.
Compared with the spherical HFB result,
the lowest energy in the axial HF calculation is higher
for $^{30}$Ne and $^{32}$Mg.
Recall that the $N=20$ magicity is broken
even in the spherical HFB for $^{30}$Ne,
while it is retained for $^{32}$Mg~\cite{ref:NS14}.
It has been established experimentally that $^{30}$Ne and $^{32}$Mg
lie on the island of inversion~\cite{ref:Ne30,ref:Mg32},
while $^{34}$Si does not~\cite{ref:Si34}.
Although the lack of the pair and rotational correlations
prohibits us from being conclusive,
the present M3Y-P6 results do not contradict
the experimental data on the shapes of these nuclei.

\begin{table}
\caption{Intrinsic mass quadrupole moment $q_0$ ($\mathrm{fm^2}$),
deformation parameter $\beta$ and energy $E$ ($\mathrm{MeV}$)
at the lowest and second lowest minima obtained in the axial HF calculations,
for the proton-deficient $N=20$ nuclei.
Results of the M3Y-P6 interaction are compared
with those obtained without $v^{(\mathrm{TN})}$ and those of the D1M interaction.
For the lowest minima,
the results obtained by minimizing $\langle H-\sum_{i<j}v_{ij}^{(\mathrm{TN})}\rangle$
are shown in the parentheses.
For reference,
the energy obtained from the spherical HFB calculation with M3Y-P6 (5th column)
and the measured energy of the ground state (right most column)~\cite{ref:AME12}
are also presented.
\label{tab:min-N20}}
\begin{ruledtabular}
\begin{tabular}{crrrrrrrrrrr}
nuclide & \multicolumn{4}{c}{M3Y-P6} & \multicolumn{3}{c}{w/o $v^{(\mathrm{TN})}$}
& \multicolumn{3}{c}{D1M} & Exp.~~\\
& $q_0$~ & $\beta$~~ & $E$~~~ & $E_\mathrm{HFB}$~ & $q_0$~ & $\beta$~~ & $E$~~~
& $q_0$~ & $\beta$~~ & $E$~~~ & $E_{g.s.}$~~\\
\hline
$^{30}$Ne & $158$ & $0.57$ & $-204.7$ & $-206.2$ & $150$ & $0.55$ & $-206.3$
 & $-22$ & $-0.08$ & $-207.3$ & $-212.3$ \vspace*{-2mm}\\
 &&&&& (~$153$ & $0.56$ & $-206.5$)\!\!&&&& \\
 & $-23$ & $-0.08$ & $-203.9$ && $27$ & $0.10$ & $-204.3$
 & $22$ & $0.08$ & $-207.2$ &\\
$^{32}$Mg & $37$ & $0.12$ & $-241.5$ & $-243.4$ & $151$ & $0.49$ & $-244.3$
 & $33$ & $0.11$ & $-245.4$ & $-249.7$ \vspace*{-2mm}\\
 &&&&& (~$159$ & $0.52$ & $-244.6$)\!\!&&&& \\
 & $168$ & $0.55$ & $-241.1$ && $240$ & $0.78$ & $-244.1$
 & $160$ & $0.52$ & $-245.4$ &\\
$^{34}$Si & $0$ & $0.00$ & $-277.5$ & $-277.5$ & $0$ & $0.00$ & $-279.3$
 & $0$ & $0.00$ & $-281.3$ & $-283.4$ \vspace*{-2mm}\\
 &&&&& (~$0$ & $0.00$ & $-279.5$)\!\!&&&& \\
 & $-140$ & $-0.41$ & $-273.0$ && $-131$ & $-0.39$ & $-276.3$
 & $-127$ & $-0.37$ & $-276.5$ &\\
\end{tabular}
\end{ruledtabular}
\end{table}

\subsection{$N=28$ nuclei: $^{40}$Mg, $^{42}$Si and $^{44}$S}

Table~\ref{tab:min-N28a} shows CHF results of $^{46}$Ar and $^{48}$Ca,
which locate at and near the $\beta$ stability.
We confirm that the $N=28$ magic number is almost kept in them.
Although $q_0$ at the lowest minimum of $^{46}$Ar is not negligible,
the neutron configuration is compatible with the $N=28$ closure,
as for a minimum in $^{44}$S discussed later.

\begin{table}
\caption{Intrinsic mass quadrupole moment $q_0$ ($\mathrm{fm^2}$),
deformation parameter $\beta$ and energy $E$ ($\mathrm{MeV}$)
at the lowest and second lowest minima
obtained in the axial HF calculations,
for the $N=28$ nuclei near the $\beta$-stability.
The spherical HFB energy with M3Y-P6
and the experimental ground-state energy~\cite{ref:AME12} are also presented.
\label{tab:min-N28a}}
\begin{ruledtabular}
\begin{tabular}{crrrrrrrr}
nuclide & \multicolumn{4}{c}{M3Y-P6} & \multicolumn{3}{c}{D1M} & Exp.~~\\
& $q_0$~ & $\beta$~~ & $E$~~~ & $E_\mathrm{HFB}$~ & $q_0$~ & $\beta$~~ & $E$~~~
 & $E_{g.s.}$~~\\
\hline
$^{46}$Ar & $-109$ & $-0.19$ & $-381.0$ & $-380.4$ & $-116$ & $-0.21$ & $-383.2$
 & $-386.9$ \\
 & $51$ & $0.09$ & $-377.6$ && $51$ & $0.09$ & $-381.0$ &\\
$^{48}$Ca & $0$ & $0.00$ & $-413.8$ & $-413.8$ & $0$ & $0.00$ & $-414.6$
 & $-416.0$ \\
 & $-189$ & $-0.31$ & $-403.2$ && $-188$ & $-0.31$ & $-407.4$ &\\
\end{tabular}
\end{ruledtabular}
\end{table}

\begin{figure}
\includegraphics[scale=1.0]{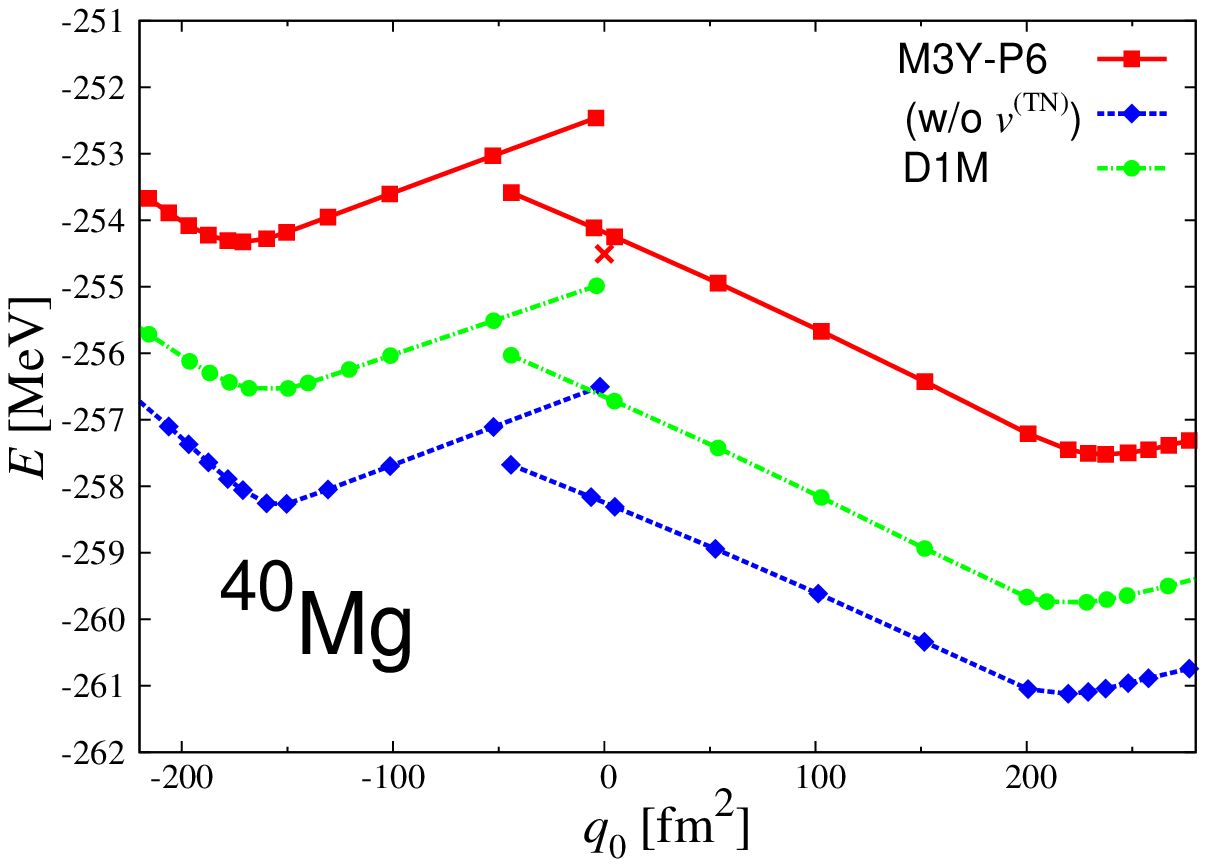}\vspace*{1cm}
\caption{(Color online) CHF results for $^{40}$Mg.
See Fig.~\protect\ref{fig:Ne30_E-q0} for conventions.
\label{fig:Mg40_E-q0}}
\end{figure}

\begin{figure}
\includegraphics[scale=1.0]{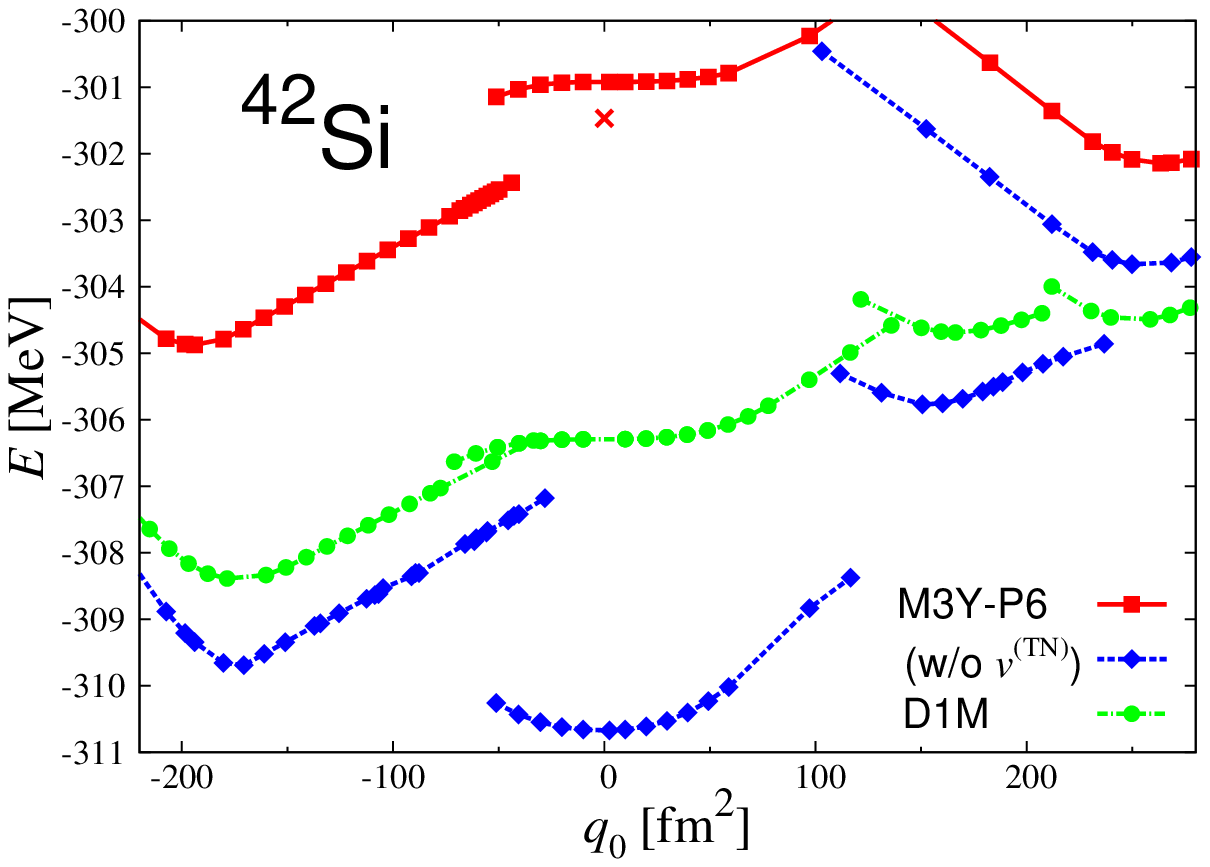}\vspace*{1cm}
\caption{(Color online) CHF results for $^{42}$Si.
See Fig.~\protect\ref{fig:Ne30_E-q0} for conventions.
\label{fig:Si42_E-q0}}
\end{figure}

\begin{figure}
\includegraphics[scale=1.0]{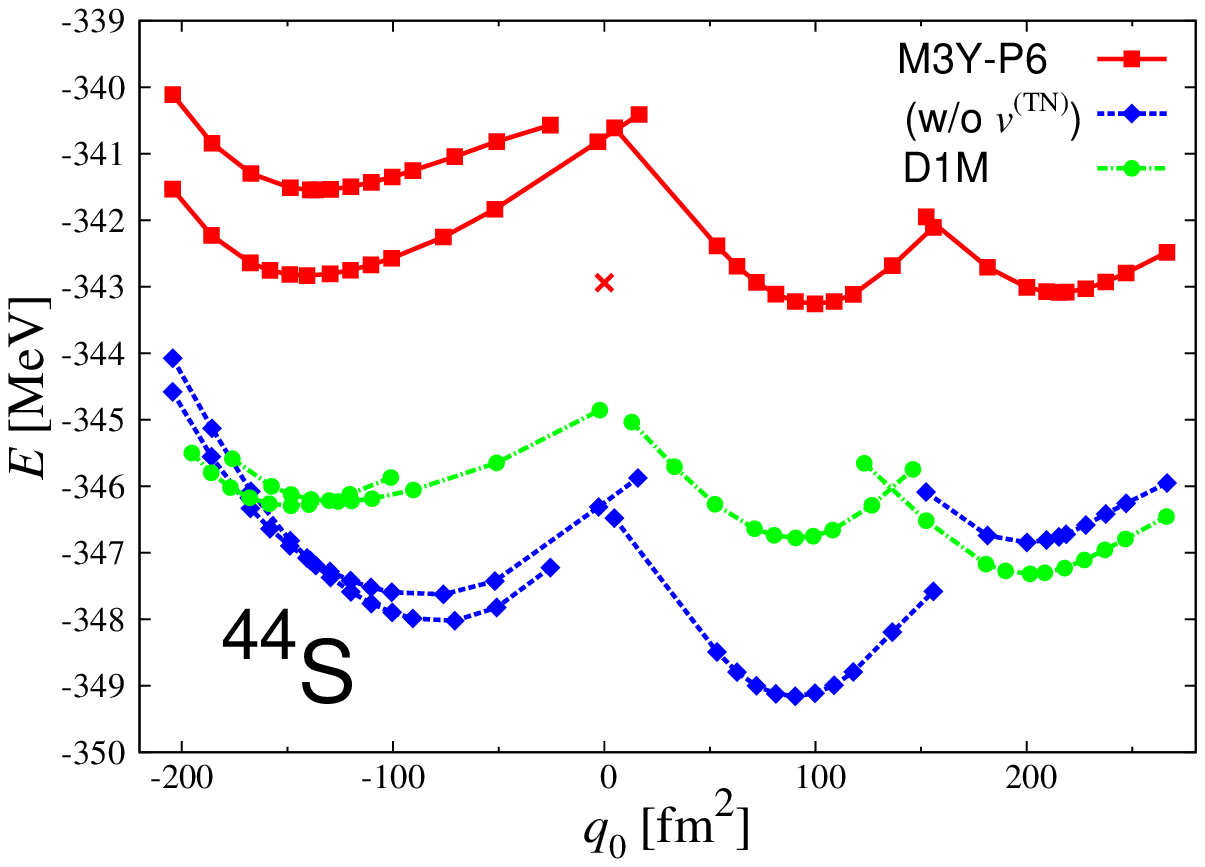}\vspace*{1cm}
\caption{(Color online) CHF results for $^{44}$S.
See Fig.~\protect\ref{fig:Ne30_E-q0} for conventions.
\label{fig:S44_E-q0}}
\end{figure}

In Figs.~\ref{fig:Mg40_E-q0}, \ref{fig:Si42_E-q0} and \ref{fig:S44_E-q0},
the energy curves for $^{40}$Mg, $^{42}$Si and $^{44}$S are presented.
The values of $q_0$ and $E$ at the lowest and second lowest minima
are listed in Table~\ref{tab:min-N28},
along with the values of $\beta$.
The spherical HFB energies and the experimental ground-state energies
are also given for reference.

\begin{table}
\caption{Intrinsic mass quadrupole moment $q_0$ ($\mathrm{fm^2}$),
deformation parameter $\beta$ and energy $E$ ($\mathrm{MeV}$)
at the lowest and second lowest minima obtained in the axial HF calculations,
for the proton-deficient $N=28$ nuclei.
The spherical HFB energy with M3Y-P6
and the experimental ground-state energy~\cite{ref:AME12} are also presented.
See Table~\ref{tab:min-N20} for legend.
\label{tab:min-N28}}
\begin{ruledtabular}
\begin{tabular}{crrrrrrrrrrr}
nuclide & \multicolumn{4}{c}{M3Y-P6} & \multicolumn{3}{c}{w/o $v^{(\mathrm{TN})}$}
& \multicolumn{3}{c}{D1M} & Exp.~~\\
& $q_0$~ & $\beta$~~ & $E$~~~ & $E_\mathrm{HFB}$~ & $q_0$~ & $\beta$~~ & $E$~~~
& $q_0$~ & $\beta$~~ & $E$~~~ & $E_{g.s.}$~~\\
\hline
$^{40}$Mg & $237$ & $0.53$ & $-257.5$ & $-254.5$ & $220$ & $0.50$ & $-261.1$
 & $220$ & $0.50$ & $-259.8$ & --- \vspace*{-2mm}\\
 &&&&& (~$222$ & $0.50$ & $-261.6$)\!\!&&&& \\
 & $-171$ & $-0.38$ & $-254.3$ && $-150$ & $-0.34$ & $-258.3$
 & $-159$ & $-0.36$ & $-256.6$ &\\
$^{42}$Si & $-194$ & $-0.40$ & $-304.9$ & $-301.5$ & $0$ & $0.00$ & $-310.7$
 & $-173$ & $-0.36$ & $-308.4$ & --- \vspace*{-2mm}\\
 &&&&& (~$0$ & $0.00$ & $-311.0$)\!\!&&&& \\
 & $263$ & $0.55$ & $-302.1$ && $-171$ & $-0.35$ & $-309.7$
 & \multicolumn{3}{c}{(not well identified)} &\\
$^{44}$S & $100$ & $0.19$ & $-343.3$ & $-342.9$ & $90$ & $0.17$ & $-349.2$
 & $202$ & $0.39$ & $-347.3$ & $-351.8$ \vspace*{-2mm}\\
 &&&&& (~~$84$ & $0.16$ & $-349.8$)\!\!&&&& \\
 & $215$ & $0.41$ & $-343.1$ && $-71$ & $-0.14$ & $-348.0$
 & $91$ & $0.17$ & $-346.8$ &\vspace*{-2mm}\\
\end{tabular}
\end{ruledtabular}
\end{table}

In $^{40}$Mg we observe a prolate minimum and an oblate minimum.
The prolate minimum is lower than the oblate minimum
and the spherical HFB energy.
This result is harmonious with several previous calculations
that do not take account of the tensor force explicitly~\cite{ref:Ter97,ref:Nak08,ref:Li11}.
Since D1M and M3Y-P6, with and without $v^{(\mathrm{TN})}$,
give almost parallel energy curves,
the $q_0$ values at the minima and the energy difference between the two minima
are insensitive to the interactions under consideration.
The tensor force yields an almost constant shift of energy,
nearly independent of $q_0$.

Energy curves in $^{42}$Si are rather complicated.
The absolute minimum has an oblate shape both in the D1M and M3Y-P6 results,
as in several previous studies~\cite{ref:Ter97,ref:Peru00,ref:Li11}.
Moreover, this minimum has a substantially lower energy
than the spherical HFB result.
As we shall argue later in detail,
the content of the occupied levels appreciably differs
from that of the $N=28$ magicity at this minimum.
Thus deformation is unambiguously predicted for this nucleus.
Although the spherical configuration does not produce
a well-developed local minimum in the D1M and M3Y-P6 results,
it produces a minimum if the tensor force is omitted
from the M3Y-P6 result, and is even lower than the oblate minimum.
It can be said that the tensor force weakens the sphericity of $^{42}$Si,
and this effect is incorporated in an effective manner in D1M.
A clear local minimum is observed on the prolate side with M3Y-P6,
which has a lower energy than the spherical configuration.
In contrast, the prolate minima with D1M have higher energies
than the spherical configuration.

There are two prolate minima in $^{44}$S.
While the $N=28$ magicity is not significantly broken
at the $q_0\approx 100\,\mathrm{fm}^2$ minimum,
it is clearly broken at the $q_0\approx 210\,\mathrm{fm}^2$ minimum.
The energies of these two minima are close to each other
both in the D1M and the M3Y-P6 results,
although the smaller (larger) $q_0$ state is lowest in the case of M3Y-P6 (D1M).
These minima have slightly lower energies than that for the spherical HFB.
Nevertheless, combined with the configuration shown in Sec.~\ref{sec:s.p.level}
and the weak deformation of the $q_0\approx 100\,\mathrm{fm}^2$ minimum,
there remains a room for an interpretation that $N=28$ remains submagic
in this nucleus.
Without the tensor force, the smaller $q_0$ configuration gives
distinctly lower energy than the larger $q_0$ one.
On the oblate side, we observe two different states with similar $q_0$,
which are distinguished by the proton configurations.
In the results of D1M and M3Y-P6 without the tensor force,
these two states have quite similar energies.
The tensor force lifts this approximate degeneracy.
Several minima on the prolate and oblate sides with close energies
suggest shape coexistence or a possibility of triaxial deformation
in this nucleus.

It is noteworthy that the shape change from nearly spherical at $^{44}$S
to oblate at $^{42}$Si to prolate at $^{40}$Mg
seems consistent with experiments~\cite{ref:Cra14},
although the pair and rotational correlations are ignored
in the present work.

\section{Single-particle levels\label{sec:s.p.level}}

Properties of the individual minima can be analyzed
in terms of the single-particle configurations,
as already referred to in the previous section.
An axially deformed s.p. level $\nu$ has the quantum number $\Omega^\pi$,
where $\Omega$ expresses the absolute value of the magnetic quantum number
along the intrinsic $z$-axis, and $\pi$ is the parity.
As a result of the time-reversal symmetry,
each level has two-fold degeneracy at any $q_0$,
known as the Kramers degeneracy.
The s.p. levels can be expanded by the spherical orbits
having the same parity and $j\geq\Omega$.
Around the $N=20$ and $28$ shell gaps in the spherical limit,
there are $n1s_{1/2}$, $n0d_{3/2}$, $n0f_{7/2}$ and $n1p_{3/2}$ orbits.
Therefore the s.p. levels around the Fermi energy
are mainly comprised of these spherical orbits.
Table~\ref{tab:Nilsson} lists an approximate correspondence
between the axial s.p. levels and the spherical orbits.
As tabulated in Table~\ref{tab:Nilsson},
the $1s_{1/2}$, $0d_{3/2}$, $0f_{7/2}$ and $1p_{3/2}$ orbits constitute
the main portion of the $\Omega^\pi_k=(1/2)^+_3$,
$\Omega^\pi_k=(1/2)^+_4,(3/2)^+_2$,
$\Omega^\pi_k=(1/2)^-_3,(3/2)^-_2,(5/2)^-_1,(7/2)^-_1$
and $\Omega^\pi_k=(1/2)^-_4,(3/2)^-_3$ levels,
where $k$ is the sequential number for each $\Omega^\pi$,
although the $(1/2)^+$ levels originating from $1s_{1/2}$ and $0d_{3/2}$
are inverted in some nuclei.
Crossing of these levels, which influences their occupation,
is relevant to the persistence and breaking of the $N=20$ and $28$ magicity.

In representing the axially deformed s.p. levels,
it is customary to use Nilsson's asymptotic quantum numbers.
Together with the label $\Omega^\pi_k$,
the Nilsson representation is given in Table~\ref{tab:Nilsson}.
However, Nilsson's quantum numbers are not always convenient
for denoting the s.p. levels in connection with the spherical limit,
because composition of the spherical orbits in the deformed s.p. level changes
as $\beta$ (or $q_0$) varies, leading to obscurity.
For this reason, we shall use the expression $\Omega^\pi_k$
instead of the Nilsson representation.
In the figures below,
the s.p. levels only at the local minima will be shown to avoid complication,
without using the CHF results.

\begin{table}
\caption{Approximate correspondence of axial s.p. levels
to spherical orbits.
The expression $\Omega^\pi_k$
and the Nilsson asymptotic quantum numbers
(``$[N\,n_3\,\Lambda\,\Omega]$'')~\cite{ref:BM2}
are presented at the left and the right of the slashes.
Components of the spherical orbits having the same $\Omega^\pi$
should mix in the axial levels,
depending on $\beta$.
The Nilsson numbers for same $\Omega^\pi$ may be inverted.
\label{tab:Nilsson}}
\begin{ruledtabular}
\begin{tabular}{ccccc}
spherical & \multicolumn{4}{c}{$\Omega^\pi_k$\,\big
 /\,$[N\,n_3\,\Lambda\,\Omega]$} \\
\hline
$0d_{5/2}$ & $(1/2)^+_2$\,\big/\,$[2\,2\,0\,\frac{1}{2}]$
 & $(3/2)^+_1$\,\big/\,$[2\,1\,1\,\frac{3}{2}]$
 & $(5/2)^+_1$\,\big/\,$[2\,0\,2\,\frac{5}{2}]$ &\\
$1s_{1/2}$ & $(1/2)^+_3$\,\big/\,$[2\,0\,0\,\frac{1}{2}]$ &&&\\
$0d_{3/2}$ & $(1/2)^+_4$\,\big/\,$[2\,1\,1\,\frac{1}{2}]$
 & $(3/2)^+_2$\,\big/\,$[2\,0\,2\,\frac{3}{2}]$ &&\\
\hline
$0f_{7/2}$ & $(1/2)^-_3$\,\big/\,$[3\,3\,0\,\frac{1}{2}]$
 & $(3/2)^-_2$\,\big/\,$[3\,2\,1\,\frac{3}{2}]$
 & $(5/2)^-_1$\,\big/\,$[3\,1\,2\,\frac{5}{2}]$
 & $(7/2)^-_1$\,\big/\,$[3\,0\,3\,\frac{7}{2}]$ \\
$1p_{3/2}$ & $(1/2)^-_4$\,\big/\,$[3\,1\,0\,\frac{1}{2}]$
 & $(3/2)^-_3$\,\big/\,$[3\,0\,1\,\frac{3}{2}]$ &&\\
\end{tabular}
\end{ruledtabular}
\end{table}

The contribution of the tensor force to the energy of the s.p. level $\nu$
is evaluated by
\begin{equation}
 \varepsilon^{(\mathrm{TN})}(\nu)
 = 2\sum_{\nu'\,(>0)} n_{\nu'}\langle\nu\nu'|v^{(\mathrm{TN})}|\nu\nu'\rangle\,,
 \label{eq:eTN}
\end{equation}
where $n_\nu$ denotes the occupation probability of $\nu$.
The sum over $\nu'$ counts only one of the members of the Kramers degeneracy,
leading to the overall coefficient of $2$.
Equation~(\ref{eq:eTN}) is a straightforward generalization
of the spherically symmetric case in Ref.~\cite{ref:NS14}.
$\varepsilon(\nu)-\varepsilon^{(\mathrm{TN})}(\nu)$ as well as $\varepsilon(\nu)$
will be plotted.
Since $v^{(\mathrm{TN})}$ is a two-body interaction,
we have $E^{(\mathrm{TN})}=\sum_{\nu\,(>0)} n_\nu\,\varepsilon^{(\mathrm{TN})}(\nu)$.

\subsection{$^{30}$Ne}

The s.p. levels of $^{30}$Ne obtained from the HF calculations with M3Y-P6,
corresponding to the minima presented in Fig.~\ref{fig:Ne30_E-q0},
are shown in Fig.~\ref{fig:Ne30_spe}.
In all of the minima,
the proton configuration is dominated by $(0d_{5/2})^2$
in terms of the spherical orbits.
At $q_0=0$ we also plot the s.p. levels
obtained from the spherical HF calculation,
in which each substate of $p0d_{5/2}$ is assumed to be equally populated.

\newlength{\spfig}  
\setlength{\spfig}{9cm}
\begin{figure}
\hspace*{-1cm}\includegraphics[height=\spfig]{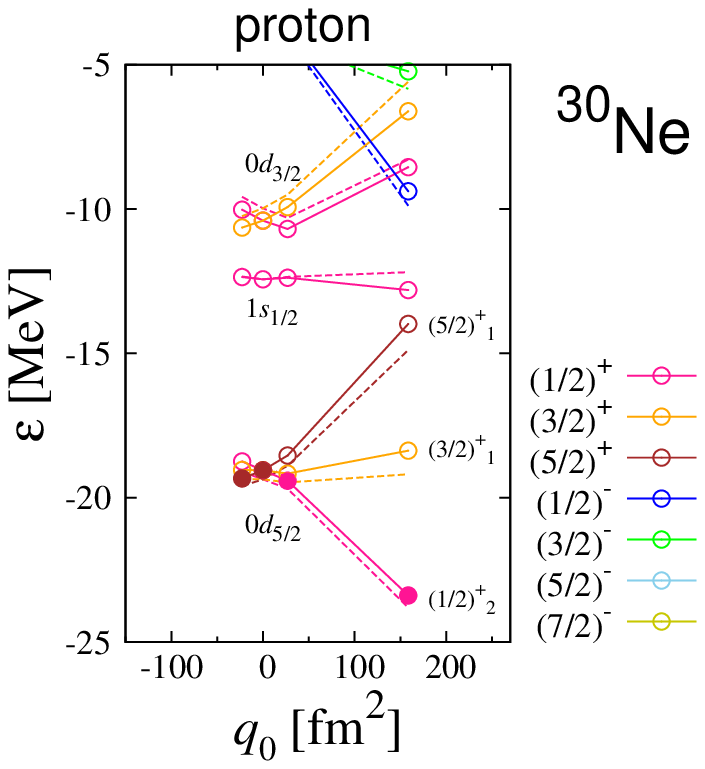}
\hspace*{-0.13\spfig}\includegraphics[height=\spfig]{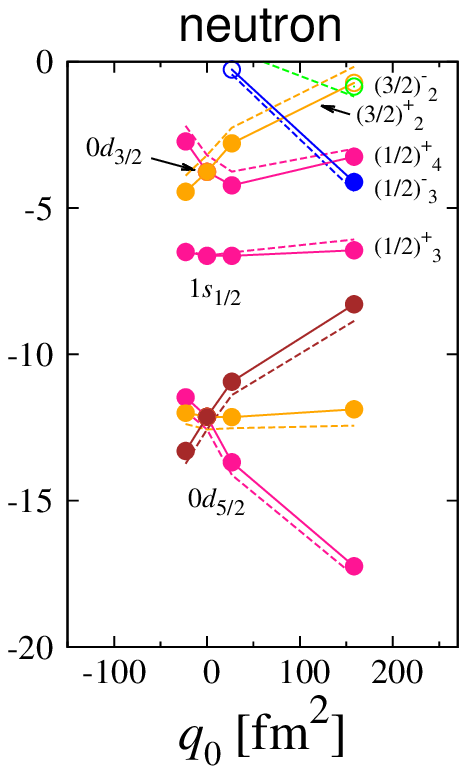}\hspace*{-4cm}
\caption{(Color online) Proton and neutron s.p. levels $\varepsilon(\nu)$
in $^{30}$Ne obtained from the HF calculations with M3Y-P6,
at the minima shown in Fig.~\ref{fig:Ne30_E-q0}.
Occupied (unoccupied) levels are represented by the filled (open) circles,
and the corresponding levels are connected by the solid lines.
The quantum number of each level $\Omega^\pi$ is indicated in the middle,
and the $\Omega^\pi_k$ label is presented for several levels
around the Fermi energy.
The dashed lines show the s.p. energies
after subtracting the tensor-force contribution,
\textit{i.e.}, $\varepsilon(\nu)-\varepsilon^{(\mathrm{TN})}(\nu)$.
The levels at $q_0=0$ are the results of the spherical HF calculation.
The labels for the spherical orbits are also attached.
\label{fig:Ne30_spe}}
\end{figure}

The configuration giving the $N=20$ magic number is broken
when the excitation from $n0d_{3/2}$
(the highest occupied level in the spherical limit)
to $n0f_{7/2}$ (the lowest unoccupied level in the spherical limit) occurs.
If the shell gap at the sphericity
(\textit{i.e.}, the energy difference between $n0d_{3/2}$ and $n0f_{7/2}$)
is compared among $^{30}$Ne, $^{32}$Mg and $^{34}$Si,
with results for the latter two
shown in Figs.~\ref{fig:Mg32_spe} and \ref{fig:Si34_spe},
it is found that the gap decreases as $Z$ decreases from $^{34}$Si to $^{30}$Ne.
This variation in the shell gap is somewhat responsible
for the quadrupole deformation
and therefore the loss of $N=20$ magicity in the proton-deficient nuclei.
The tensor force enhances this trend, though not being dominant.

At the two minima with $|q_0|<50\,\mathrm{fm}^2$ in $^{30}$Ne,
the neutron configuration well matches $(0d_{5/2})^6(1s_{1/2})^2(0d_{3/2})^4$,
\textit{i.e.}, the configuration giving the $N=20$ closure.
This indicates that the $N=20$ magicity is hardly broken at these minima.
In contrast, the $n(1/2)^-_3$ level is occupied at the minimum
with $q_0\approx 160\,\mathrm{fm}^2$,
because of its crossing with $n(3/2)^+_2$.
This clearly indicates that the $N=20$ magicity is broken in this state,
which gives the absolute minimum within the HF approximation
as shown in Fig.~\ref{fig:Ne30_E-q0}.
It is noted that the HF states before and after the level crossing
are mutually orthogonal,
because the quantum numbers of the occupied s.p. levels do not match.

The tensor force shifts up and down individual s.p. levels,
as recognized from $\varepsilon-\varepsilon^{(\mathrm{TN})}$
represented by the dashed lines in Fig.~\ref{fig:Ne30_spe}.
It also affects the slopes of the levels against $q_0$,
but does not drastically change them.
It is convenient to describe the tensor-force effects
in terms of the s.p. orbits in the spherical limit.
The tensor force operates repulsively (attractively)
on neutron $j=\ell+1/2$ ($j=\ell-1/2$) orbits
as a proton $j=\ell+1/2$ orbit is occupied~\cite{ref:Vtn}.
The tensor-force contribution almost vanishes
if summed over the s.p. levels belonging to an $\ell s$ closed shell,
as proven in Appendix.
Since $p0d_{5/2}$ is partially occupied in $^{30}$Ne,
the $n0f_{7/2}$ level shifts up while $n0d_{3/2}$ shifts down.
Thus the tensor force tends to delay the level crossing
and therefore the breakdown of the $N=20$ magicity.
This is exemplified by the crossing between $n(3/2)^+_2$ and $n(1/2)^-_3$
in the right panel of Fig.~\ref{fig:Ne30_spe},
although it does not change the situation
that the deformed configuration gives the absolute minimum of $^{30}$Ne
within the HF framework.

\subsection{$^{32}$Mg}

\begin{figure}
\hspace*{-1cm}\includegraphics[height=\spfig]{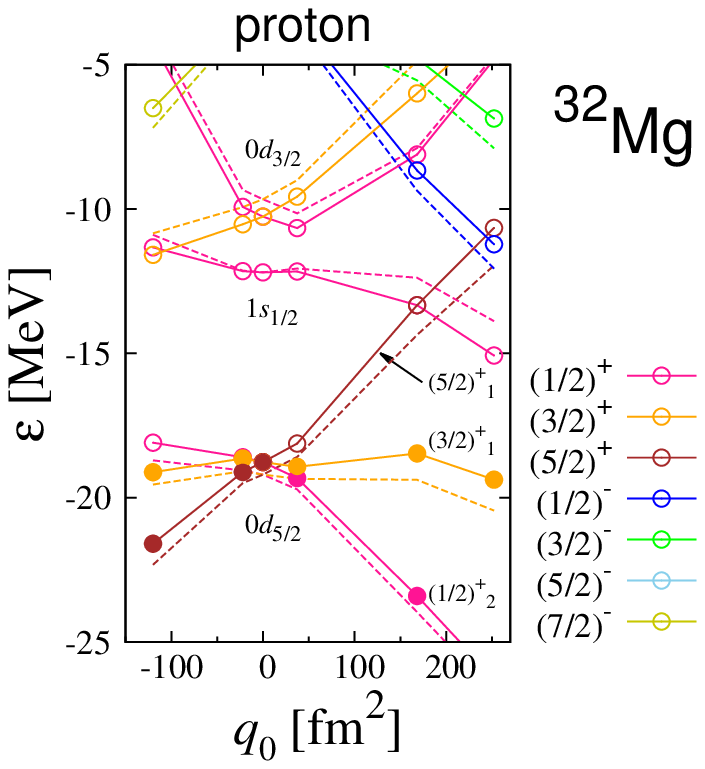}
\hspace*{-0.13\spfig}\includegraphics[height=\spfig]{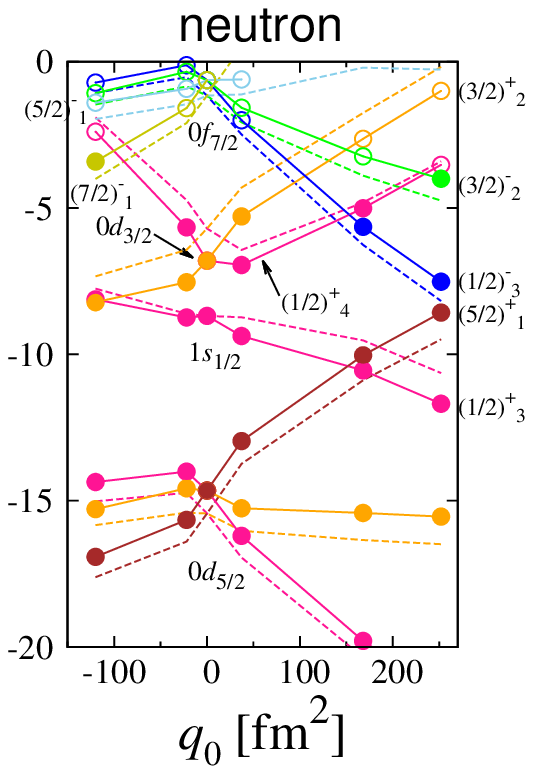}\hspace*{-4cm}
\caption{(Color online) Proton and neutron s.p. levels in $^{32}$Mg
obtained from the HF calculations with M3Y-P6,
at the minima shown in Fig.~\ref{fig:Mg32_E-q0}.
See Fig.~\ref{fig:Ne30_spe} for conventions.
\label{fig:Mg32_spe}}
\end{figure}

The s.p. levels of $^{32}$Mg obtained with M3Y-P6
are depicted in Fig.~\ref{fig:Mg32_spe},
at the minima shown in Fig.~\ref{fig:Mg32_E-q0}.
For the range of $q_0$ shown here,
the proton configuration is interpreted as $(0d_{5/2})^4$.
While the neutron configuration is compatible with the $N=20$ closure
for the two minima with $|q_0|<50\,\mathrm{fm}^2$,
the $N=20$ magicity is obviously broken at the other minima
with higher $|q_0|$, which involve excitations to $n0f_{7/2}$ .

The $q_0\approx 170\,\mathrm{fm}^2$ minimum,
which arises after the crossing of $n(3/2)^+_2$ and $n(1/2)^-_3$,
has a similar neutron configuration
to that of the well-deformed minimum in $^{30}$Ne.
At the $q_0\approx 250\,\mathrm{fm}^2$ minimum,
the $n(3/2)^-_2$ level is also occupied instead of $n(1/2)^+_4$.
At the $q_0\approx -120\,\mathrm{fm}^2$ minimum,
the $n(7/2)^-_1$ level is occupied instead of $n(1/2)^+_4$.

Compared with $\varepsilon-\varepsilon^{(\mathrm{TN})}$,
the s.p. energy $\varepsilon$ of $n0d_{3/2}$ is lower
while that of $n0f_{7/2}$ is higher,
owing to the occupation of $p0d_{5/2}$.
This makes the energies of the large $|q_0|$ configurations relatively high,
accounting for the tensor-force effect in Fig.~\ref{fig:Mg32_E-q0}.
Still the energy of the $q_0\approx 170\,\mathrm{fm}^2$ state
competes with those of the $|q_0|< 50\,\mathrm{fm}^2$ states.

\subsection{$^{34}$Si}

\begin{figure}
\hspace*{-1cm}\includegraphics[height=\spfig]{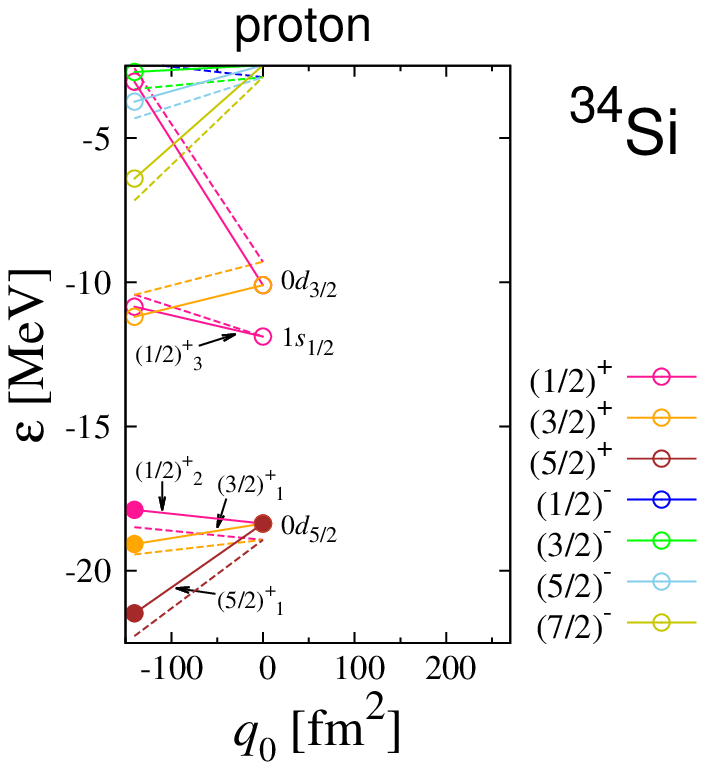}
\hspace*{-0.13\spfig}\includegraphics[height=\spfig]{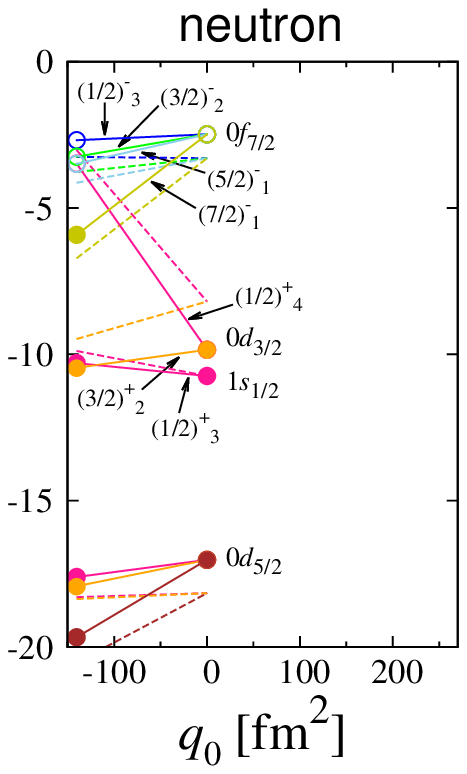}\hspace*{-4cm}
\caption{(Color online) Proton and neutron s.p. levels in $^{34}$Si
obtained from the HF calculations with M3Y-P6,
at the minima shown in Fig.~\ref{fig:Si34_E-q0}.
See Fig.~\ref{fig:Ne30_spe} for conventions.
\label{fig:Si34_spe}}
\end{figure}

In Fig.~\ref{fig:Si34_spe} we display the s.p. levels
obtained for $^{34}$Si with M3Y-P6,
at the minima presented in Fig.~\ref{fig:Si34_E-q0}.
For this nucleus, a spherically symmetric solution is obtained
within the axial HF calculations,
in which $p0d_{5/2}$ is fully occupied
and the $N=20$ magicity is strictly maintained.
This state provides the absolute minimum in Fig.~\ref{fig:Si34_E-q0}.
We confirm in Fig.~\ref{fig:Si34_spe}
that the tensor force strengthens this configuration
by increasing the shell gap between $n0d_{3/2}$ and $n0f_{7/2}$,
assisting the doubly magic nature of $^{34}$Si
as viewed in Fig.~\ref{fig:Si34_E-q0}.

\subsection{$^{40}$Mg}

In the $N=20$ nuclei
the highest occupied neutron s.p. level has even-parity in the spherical limit,
while the lowest unoccupied level has odd-parity.
Thereby the HF state after the level crossing is orthogonal
to the configuration keeping the $N=20$ closure,
and the breakdown of the $N=20$ magicity after the crossing is clear
from the parities of the occupied s.p. levels.
In contrast, both the highest occupied and the lowest unoccupied levels
have odd-parities in the $N=28$ nuclei.
Still, the states before and after the level crossing
remain orthogonal on the prolate side,
since the highest occupied level in the $N=28$ closure is
$\Omega^\pi_k=(7/2)^-_1$,
which cannot be formed by the low-lying unoccupied levels.
However, this is not the case on the oblate side,
where the $\Omega^\pi_k=(1/2)^-_3$ level is the highest among the occupied levels
near the sphericity.
More careful analysis is needed
for investigating the breakdown of the $N=28$ magic number,
particularly on the oblate side.

\begin{figure}
\hspace*{-1cm}\includegraphics[height=\spfig]{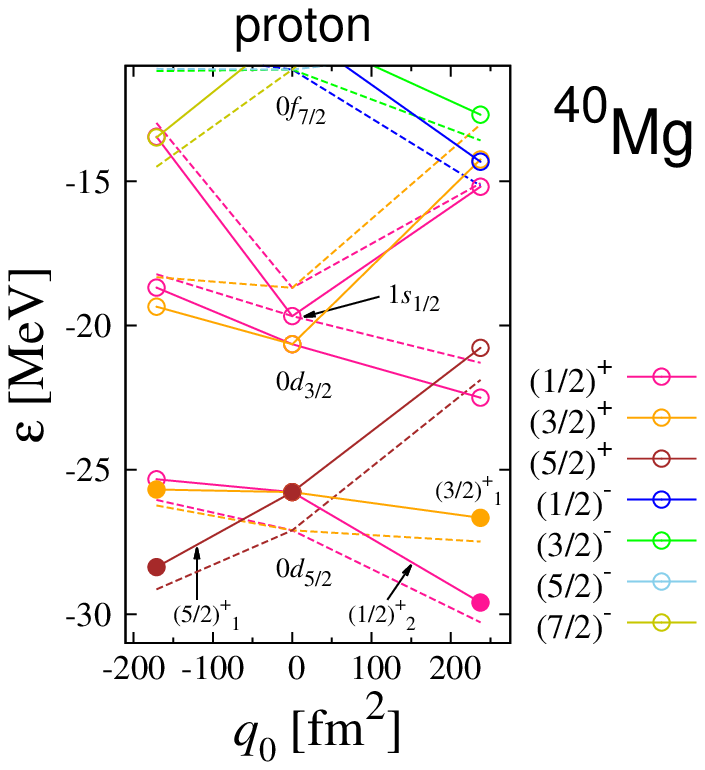}
\hspace*{-0.13\spfig}\includegraphics[height=\spfig]{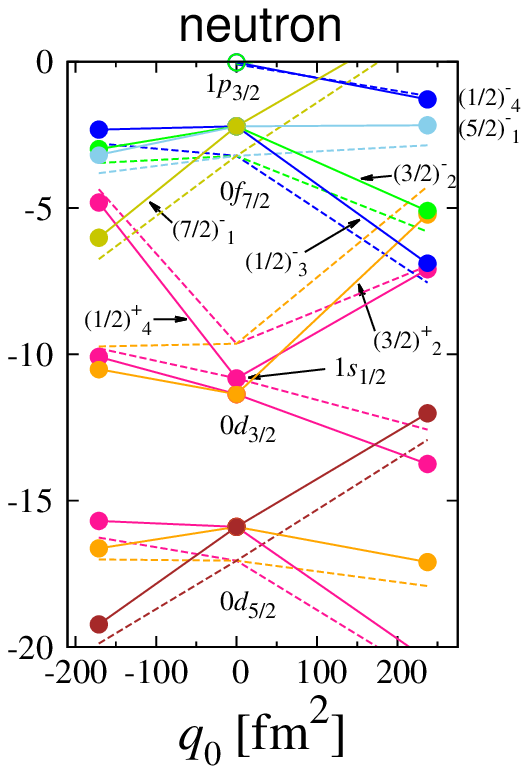}\hspace*{-4cm}
\caption{(Color online) Proton and neutron s.p. levels in $^{40}$Mg
obtained from the HF calculations with M3Y-P6,
at the minima shown in Fig.~\ref{fig:Mg40_E-q0}.
See Fig.~\ref{fig:Ne30_spe} for conventions.
\label{fig:Mg40_spe}}
\end{figure}

Comparing the $N=28$ spherical shell gaps
(\textit{i.e.}, the energy difference between $n0f_{7/2}$ and $n1p_{3/2}$)
for $^{40}$Mg, $^{42}$Si and $^{44}$S,
which can be read from the s.p. energies at $q_0=0$
in Figs.~\ref{fig:Mg40_spe}\,--\,\ref{fig:S44_spe},
we find that the gap diminishes for decreasing $Z$ from $^{44}$S to $^{40}$Mg
as in the $N=20$ case.
The tensor force makes this trend stronger.
This result was already presented in Fig.~11 of Ref.~\cite{ref:NS14}.

The s.p. levels for $^{40}$Mg are shown in Fig.~\ref{fig:Mg40_spe}.
At the $q_0\approx -170\mathrm{fm}^2$ minimum of $^{40}$Mg,
the nuclear wave function can overlap
with the configuration that maintains the $N=28$ magicity.
However, by decomposing wave functions of the occupied s.p. levels
into the spherical bases,
the neutron $\Omega^\pi_k=(3/2)^-_2$ level is found
to consist of $f_{7/2}$ only by $46\%$
while it consists of $p_{3/2}$ by $53\%$.
The $n(1/2)^-_3$ level consists of $f_{7/2}$ by $47\%$
while it consists of $p_{3/2}$ and $p_{1/2}$ by $38\%$ and $13\%$, respectively.
Such significant presence of the $p$ orbits indicates
that this state is well deformed with breakdown of the $N=28$ magicity.

At the $q_0\approx 240\,\mathrm{fm}^2$ minimum,
the $n(1/2)^-_4$ level originating from $n1p_{3/2}$ is occupied
while the $n(7/2)^-_1$ level is open.
Since this configuration is orthogonal to the one under the $N=28$ closure
as mentioned above,
the $N=28$ magicity is broken in this state,
which is the absolute minimum in the axial HF calculation.

Although it seems irrelevant to the structure of the ground state,
it is noteworthy that the tensor force gives rise to the inversion
of $0d_{3/2}$ and $1s_{1/2}$ both for protons and neutrons in the spherical limit.
The same holds for the other $N=28$ nuclei,
and a similar inversion has been pointed out for Ca isotopes~\cite{ref:NSM13}.
It may deserve investigating how this inversion affects
excitation property of these nuclei.

\subsection{$^{42}$Si}

\begin{figure}
\hspace*{-1cm}\includegraphics[height=\spfig]{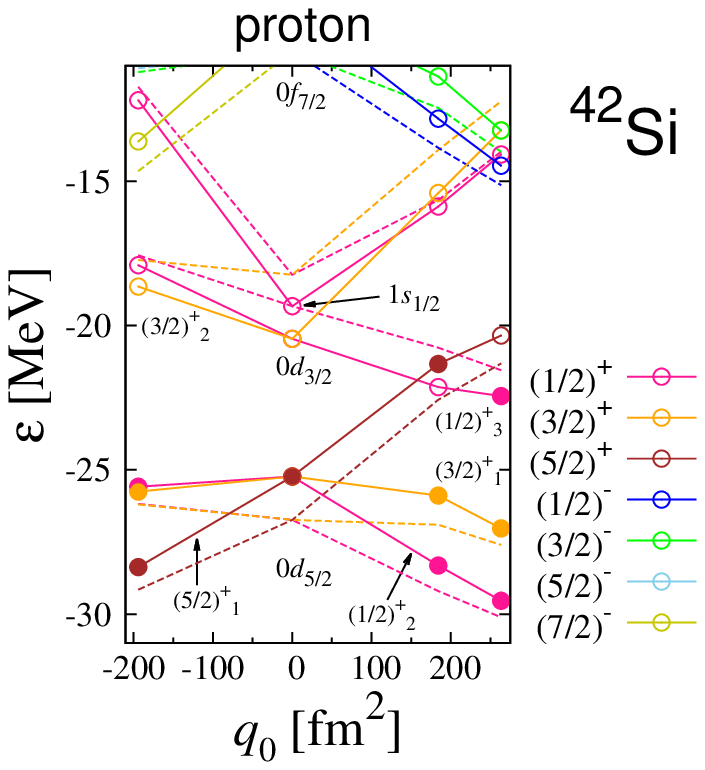}
\hspace*{-0.13\spfig}\includegraphics[height=\spfig]{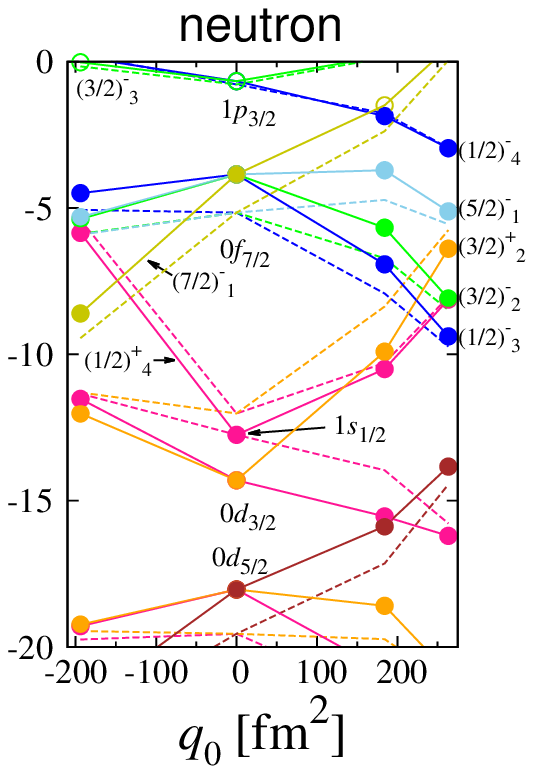}\hspace*{-4cm}
\caption{(Color online) Proton and neutron s.p. levels in $^{42}$Si
obtained from the HF calculations with M3Y-P6,
at the minima shown in Fig.~\ref{fig:Si42_E-q0}.
See Fig.~\ref{fig:Ne30_spe} for conventions.
\label{fig:Si42_spe}}
\end{figure}

Figure~\ref{fig:Si42_spe} shows the s.p. levels in $^{42}$Si.
The absolute minimum obtained in the axial HF calculation is oblate,
yielding $q_0\approx -190\,\mathrm{fm}^2$.
In this state, the occupied neutron level $\Omega^\pi_k=(3/2)^-_2$
has $44\%$ of the $f_{7/2}$ component
while $55\%$ of the $p_{3/2}$ component,
and the $n(1/2)^-_3$ level has $49\%$ of $f_{7/2}$
while $34\%$ and $15\%$ of $p_{3/2}$ and $p_{1/2}$, respectively.
Together with the large $|q_0|$ value,
this indicates that the state is well deformed
with breakdown of the $N=28$ magicity.

The $N=28$ magic number is also broken at the minima
observed on the prolate side.
This is clarified by the level crossing as in $^{40}$Mg.

\subsection{$^{44}$S}

\begin{figure}
\hspace*{-1cm}\includegraphics[height=\spfig]{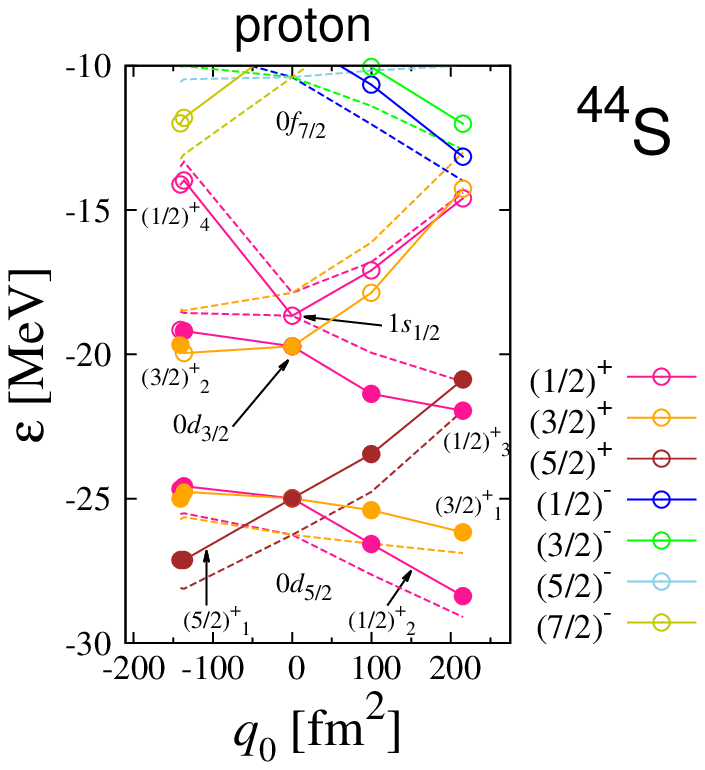}
\hspace*{-0.13\spfig}\includegraphics[height=\spfig]{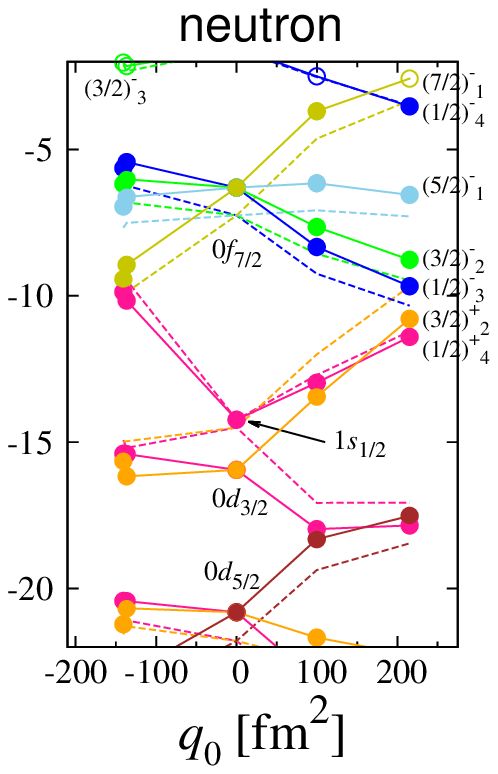}\hspace*{-4cm}
\caption{(Color online) Proton and neutron s.p. levels in $^{44}$S
obtained from the HF calculations with M3Y-P6,
at the minima shown in Fig.~\ref{fig:S44_E-q0}.
See Fig.~\ref{fig:Ne30_spe} for conventions.
\label{fig:S44_spe}}
\end{figure}

The s.p. levels for $^{44}$S are presented in Fig.~\ref{fig:S44_spe}.
Three minima compete in $^{44}$S, as has been shown in Fig.~\ref{fig:S44_E-q0}.
Despite their non-vanishing $|q_0|$,
the $N=28$ magicity is not severely broken in two of these three states.

At the oblate minimum with $q_0\approx -140\,\mathrm{fm}^2$,
both the neutron $\Omega^\pi_k=(3/2)^-_2$ and $(1/2)^-_3$ levels consist
of the $f_{7/2}$ component by $72\%$,
and the occupation probability of $nf_{7/2}$ reaches $85\%$.
There is another state with a similar $q_0$
in which the proton configuration is different.
Keep in mind that neutrons occupy $0f_{7/2}$ to a significant extent,
but not its $\ell s$ partner.
While the highest occupied proton level is $(3/2)^+_2$ in the lower state,
which is dominated by the $d_{3/2}$ component ($80\%$),
the highest level in the higher state is $(1/2)^+_3$
in which $d_{3/2}$ ($45\%$) and $s_{1/2}$ ($33\%$) mix.
Because of this difference in the contents of the spherical orbitals,
the repulsive effect of the tensor force is weaker for the lower state,
providing a sizable energy difference between these two states.
For the prolate minimum with $q_0\approx 100\,\mathrm{fm}^2$,
no level crossing occurs for neutrons
and the $\Omega^\pi_k=(3/2)^-_2$ [$(1/2)^-_3$] level
exhibits a $92\%$ ($88\%$) probability of $f_{7/2}$,
resulting in the $95\%$ occupation probability of $nf_{7/2}$.

The $N=28$ magic number is broken
at the minimum having $q_0\approx 210\,\mathrm{fm}^2$,
as is obvious from the level crossing of $n(7/2)^-_1$ and $n(1/2)^-_4$.

\subsection{Role of tensor force\label{subec:TN}}

As has been pointed out earlier,
the tensor force contributes repulsively to the total energy.
This is accounted for as follows.
Occupation of a proton $j=\ell+1/2$ orbit heightens
the energies of neutron $j=\ell+1/2$ orbits
while it lowers the energies of $j=\ell-1/2$ orbits.
If pairs of neutron $\ell s$ partners
(\textit{i.e.}, $j=\ell\pm 1/2$ orbits) are fully populated,
the contribution of the tensor force becomes negligibly small.
However, owing to the $\ell s$ splitting,
the occupation probability of a valence $j=\ell+1/2$ orbit is necessarily higher
than that of its $\ell s$ partner.
Therefore the net effect of the tensor force is repulsive unless negligible.

Deformation is enhanced
if the repulsive effect of the tensor force is weak in the deformed state
compared with the spherical state, and \textit{vice versa}.
In all nuclei under discussion,
protons occupy $0d_{5/2}$ with a significantly higher probability
than that of $0d_{3/2}$.
In the $N=28$ nuclei, the repulsive effect of the tensor force is maximal
in the spherical state where $n0f_{7/2}$ is fully occupied.
If the system is deformed, it tends toward the spin saturation,
which suppresses the tensor-force effects.
Thus the tensor force facilitates the quadrupole deformation
in the $N=28$ nuclei.
It is commented that the tensor-force effects on the deformation
in the $N=28$ nuclei have been argued within the shell model
in Ref.~\cite{ref:Utsu12}.

In the $N=20$ nuclei the tensor-force effects are small in the spherical state,
because the spin degrees of freedom (d.o.f.) for neutrons are saturated.
Once the system is deformed,
the spin saturation is somewhat broken,
and the tensor-force effects become sizable.
This mechanism works stronger after level crossing.
Because protons partially occupy $0d_{5/2}$ in this region,
the neutron $j=\ell+1/2$ ($j=\ell-1/2$) level shifts up (down)
owing to the tensor force.
If level crossing occurs,
the neutrons occupying $0d_{3/2}$ are moved to a level dominated by $0f_{7/2}$.
This causes energy loss of the deformed state.
As a result the tensor force tends to suppress the deformation
in the $N=20$ nuclei,
opposite to what happens in the $N=28$ case.

As shown in Figs.~\ref{fig:Ne30_E-q0}\,--\,\ref{fig:S44_E-q0},
some of the tensor-force effects in the M3Y-P6 results
are reproduced in the D1M results in an effective manner.
This is also consistent with the results for the $N=28$ isotones
based on the relativistic EDF with the DD-PC1 parameter-set~\cite{ref:Li11},
which does not contain the tensor force.
This indicates that certain aspects of the tensor-force effects
can be compensated by other channels.
However, the tensor-force effects yield the $Z$-dependence of the s.p. levels,
which are not well converted to the other channels~\cite{ref:NS14}.
In the energy curves in Figs.~\ref{fig:Ne30_E-q0}\,--\,\ref{fig:S44_E-q0},
we observe a visible difference between the M3Y-P6 and D1M results
(apart from constant shifts)
for the minimum at $q_0\approx 170\,\mathrm{fm}^2$ in $^{42}$Si
and the minimum at $q_0\approx -140\,\mathrm{fm}^2$ in $^{44}$S.
In addition, we have found
that M3Y-P6 favors deformation more than D1M in $^{30}$Ne.

\section{Summary\label{sec:summary}}

We have implemented
axial Hartree-Fock (including constrained Hartree-Fock) calculations
with the semi-realistic interaction M3Y-P6,
which contains a realistic tensor force.
This is the first application of the M3Y-type semi-realistic interaction
to deformed nuclei.
Effects of the tensor force on deformation have been investigated
for the neutron-rich $N=20$ and $28$ nuclei,
$^{30}$Ne, $^{32}$Mg, $^{34}$Si, $^{40}$Mg, $^{42}$Si and $^{44}$S.
With few exceptions, the tensor force does not significantly affect
the intrinsic quadrupole moments that give the energy minima,
indicating that its main effect is
to create a configuration-dependent energy shift.
It is reasonable to consider the tensor-force effects
in terms of the occupancy of spherical orbits.

In the $N=28$ nuclei, the tensor force favors deformation,
as accounted for by the full occupation
of $j=\ell+1/2$ (\textit{i.e.}, $n0f_{7/2}$) for $N=28$~\cite{ref:Sky-TNS}.
On the contrary, the tensor force tends to favor sphericity
in the $N=20$ nuclei,
because of the $\ell s$-closure of the $N=20$ magic number.
In particular, crossing between a $j=\ell-1/2$ (\textit{i.e.}, $n0d_{3/2}$) level
and a $j=\ell+1/2$ (\textit{i.e.}, $n0f_{7/2}$) level
in a deformed configuration leads to an energy loss.
While most of these tensor-force effects seem to be absorbed
into the other interaction channels,
as concluded by comparing the M3Y-P6 results to the D1M results,
a notable difference remains for several configurations.

Although inclusion of the pair and rotational correlations
should be awaited,
the present results obtained with M3Y-P6
do not contradict experimental findings.
The shape change in the $N=28$ isotones seems consistent with experiments.
For the $N=20$ nuclei, the deformed ground state of $^{30}$Ne
and the spherical ground state of $^{34}$Si match the existing data quite well.
Whereas the lowest minimum in $^{32}$Mg
is nearly spherical in the present calculation,
we have obtained another minimum with a close energy,
which suggests shape coexistence.
In a previous study,
magic numbers were well predicted in a wide range of the nuclear chart
by the spherical HFB calculations using the M3Y-P6 interaction.
As an exception, $N=20$ was not broken in $^{32}$Mg.
The present result hints that,
in the approaches based on semi-realistic interactions,
this problem might be alleviated within the MF regime
if the deformation is properly taken into account,
not invoking beyond-MF effects as in some other studies.

\begin{acknowledgments}
%
The authors are grateful to T. Inakura for discussions.
This work is financially supported in part
by the JSPS KAKENHI Grant Number~24105008 and Grant Number~16K05342.
Some of the numerical calculations have been performed on HITAC SR16000s
at Institute of Management and Information Technologies in Chiba University,
at Information Technology Center in University of Tokyo,
at Information Initiative Center in Hokkaido University,
and at Yukawa Institute for Theoretical Physics in Kyoto University.
\end{acknowledgments}

\appendix*
\section{Tensor-force effects under spin saturation}

In this Appendix,
we prove that the tensor force does not affect the s.p. energies
in spin-saturated many-body states,
and that the contribution of the tensor force to the s.p. energies vanishes
if it is summed so that the spin d.o.f. is frozen.
The latter is a generalization of Eq.~(4) in Ref.~\cite{ref:Vtn}
for spherical s.p. orbits.

We consider the following interaction between $i$-th and $j$-th nucleons,
\begin{equation}
 v_{ij}^{(\mathrm{X})} = F^{(\lambda)}(\mathbf{r}_i-\mathbf{r}_j)\cdot
 [s_i s_j]^{(\lambda)}
 = \sum_\mu (-)^\mu\,F^{(\lambda)}_{-\mu}(\mathbf{r}_i-\mathbf{r}_j)\,
 [s_i s_j]^{(\lambda)}_\mu\,, \label{eq:Vx}
\end{equation}
where $\lambda$ is the rank under the rotation, $\mu$ is the $z$-component,
$F^{(\lambda)}$ is a proper function,
$s$ is the spin operator and $[~]^{(\lambda)}$ represents the tensor coupling.
The index for the isospin d.o.f. is dropped,
because it is not essential for the arguments here.
Note that we have $\lambda=2$ for the tensor force;
\textit{i.e.}, $\mathrm{X}=\mathrm{TN}$.
$F^{(\lambda)}$ can be decomposed as
\begin{equation}
 F^{(\lambda)}_{-\mu}(\mathbf{r}_i-\mathbf{r}_j) = \sum_\alpha
 f_{\lambda\mu;\alpha}(\mathbf{r}_i)\,\bar{f}_{\lambda\mu;\alpha}(\mathbf{r}_j)\,,
 \label{eq:F}
\end{equation}
by introducing appropriate functions $f_{\lambda\mu;\alpha}$
and $\bar{f}_{\lambda\mu;\alpha}$.
For instance, such a presentation can be obtained
using the Fourier transform (see Eq.~(21) in Ref.~\cite{ref:NS02});
then $\alpha$ reads the momentum.
$v^{(\mathrm{X})}$ is now separated as a product of s.p. operators,
at the expense of the sum over $\alpha$,
\begin{equation}
 v_{ij}^{(\mathrm{X})} = \sum_{\mu\mu_1\mu_2} (-)^\mu\,(1\,\mu_1\,1\,\mu_2|\lambda\,\mu)
 \sum_\alpha \big[f_{\lambda\mu;\alpha}(\mathbf{r}_i)\,s_{i,\mu_1}\big]\,
 \big[\bar{f}_{\lambda\mu;\alpha}(\mathbf{r}_j)\,s_{j,\mu_2}\big]\,. \label{eq:Vx2}
\end{equation}

Let us consider the contribution of $v^{(\mathrm{X})}$ to the s.p. energy
as in Eq.~(\ref{eq:eTN}),
\begin{equation}
 \varepsilon^{(\mathrm{X})}(\nu)
 = \sum_{\nu_1\,(\leq\varepsilon_\mathrm{F})}
 \big[\langle\nu\nu_1|v_{ij}^{(\mathrm{X})}|\nu\nu_1\rangle'
 - \langle\nu\nu_1|v_{ij}^{(\mathrm{X})}|\nu_1\nu\rangle'\big]\,.
 \label{eq:eX}
\end{equation}
Here the sum $\sum_{\nu_1\,(\leq\varepsilon_\mathrm{F})}$ runs
over all of the occupied s.p. levels,
and $\langle~|~|~\rangle'$ indicates a non-anti-symmetrized matrix element.
Owing to Eq.~(\ref{eq:Vx2}), we separate the two-body matrix elements
into products of one-body matrix elements,
and drop the indices $i$ and $j$, obtaining
\begin{equation}\begin{split}
 \varepsilon^{(\mathrm{X})}(\nu) &= \sum_{\mu\mu_1\mu_2}
 (-)^\mu\,(1\,\mu_1\,1\,\mu_2|\lambda\,\mu)
 \sum_\alpha \sum_{\nu_1\,(\leq\varepsilon_\mathrm{F})}
 \big[\langle\nu|f_{\lambda\mu;\alpha}(\mathbf{r})\,s_{\mu_1}|\nu\rangle\,
 \langle\nu_1|\bar{f}_{\lambda\mu;\alpha}(\mathbf{r})\,s_{\mu_2}|\nu_1\rangle \\
 &\mbox{\hspace*{6.2cm}}
 - \langle\nu|f_{\lambda\mu;\alpha}(\mathbf{r})\,s_{\mu_1}|\nu_1\rangle\,
 \langle\nu_1|\bar{f}_{\lambda\mu;\alpha}(\mathbf{r})\,s_{\mu_2}|\nu\rangle\big]\,.
\end{split}\label{eq:eX2}\end{equation}
As long as we handle a many-body state that retains the time-reversal symmetry,
the direct term in Eq.~(\ref{eq:eX2}) vanishes
(see Appendix~A in Ref.~\cite{ref:VB72}) because
\begin{equation}
 \sum_{\nu_1\,(\leq\varepsilon_\mathrm{F})}
 \langle\nu_1|\bar{f}_{\lambda\mu;\alpha}(\mathbf{r})\,s_{\mu_2}|\nu_1\rangle =0\,,
\end{equation}
yielding
\begin{equation}
 \varepsilon^{(\mathrm{X})}(\nu) = -\sum_{\mu\mu_1\mu_2}
 (-)^\mu\,(1\,\mu_1\,1\,\mu_2|\lambda\,\mu)
 \sum_\alpha \sum_{\nu_1\,(\leq\varepsilon_\mathrm{F})}
 \langle\nu|f_{\lambda\mu;\alpha}(\mathbf{r})\,s_{\mu_1}|\nu_1\rangle\,
 \langle\nu_1|\bar{f}_{\lambda\mu;\alpha}(\mathbf{r})\,s_{\mu_2}|\nu\rangle\,.
\label{eq:eX3}\end{equation}
It should be noticed that
$\sum_{\nu_1\,(\leq\varepsilon_\mathrm{F})}|\nu_1\rangle\,\langle\nu_1|$
can be regarded as a one-body operator.

If the spin d.o.f. saturate in the many-body state,
the one-body operator
$\sum_{\nu_1\,(\leq\varepsilon_\mathrm{F})}|\nu_1\rangle\,\langle\nu_1|$
behaves as a scalar in the spin space
and commutes with the spin operator $s_\mu$.
Therefore we can rewrite Eq.~(\ref{eq:eX3}) as
\begin{equation}\begin{split}
 \varepsilon^{(\mathrm{X})}(\nu) &= -\sum_{\mu\mu_1\mu_2}
 (-)^\mu\,(1\,\mu_1\,1\,\mu_2|\lambda\,\mu)
 \sum_\alpha \sum_{\nu_1\,(\leq\varepsilon_\mathrm{F})}
 \langle\nu|f_{\lambda\mu;\alpha}(\mathbf{r})|\nu_1\rangle\,
 \langle\nu_1|\bar{f}_{\lambda\mu;\alpha}(\mathbf{r})\,s_{\mu_1}s_{\mu_2}|\nu\rangle \\
 &= -\sum_\mu (-)^\mu \sum_\alpha \sum_{\nu_1\,(\leq\varepsilon_\mathrm{F})}
 \langle\nu|f_{\lambda\mu;\alpha}(\mathbf{r})|\nu_1\rangle\,
 \langle\nu_1|\bar{f}_{\lambda\mu;\alpha}(\mathbf{r})\,
 [s\,s]^{(\lambda)}_\mu|\nu\rangle\,.
\end{split}\label{eq:eX3-ss}\end{equation}
However, we have
\begin{equation}
 [s\,s]^{(\lambda)}_\mu = \left\{\begin{array}{cl}
 -\frac{\sqrt{3}}{4} &\mbox{(for $\lambda=\mu=0$)}\\
 -\frac{1}{\sqrt{2}}s_\mu &\mbox{(for $\lambda=1$)}\\
 0 &\mbox{(for $\lambda=2$)}\end{array}\right.\,.
\label{eq:ss}\end{equation}
Thus, $\varepsilon^{(\mathrm{TN})}(\nu)=0$ follows
because $\lambda=2$ for $\mathrm{X}=\mathrm{TN}$.

If $\varepsilon^{(\mathrm{X})}(\nu)$ is summed over the s.p. states
so that the spin d.o.f. is frozen (as for $\ell s$ partners),
for which the summation is expressed by $\sum^\sigma$,
similar arguments are applicable
even when the spin d.o.f. do not saturate in the many-body state.
Since $\sum_\nu^\sigma|\nu\rangle\,\langle\nu|$ is scalar in the spin space,
we have
\begin{equation}\begin{split}
 {\sum_\nu}^\sigma\varepsilon^{(\mathrm{X})}(\nu) &= -{\sum_\nu}^\sigma\sum_{\mu\mu_1\mu_2}
 (-)^\mu\,(1\,\mu_1\,1\,\mu_2|\lambda\,\mu)
 \sum_\alpha \sum_{\nu_1\,(\leq\varepsilon_\mathrm{F})}
 \langle\nu|f_{\lambda\mu;\alpha}(\mathbf{r})\,s_{\mu_1}|\nu_1\rangle\,
 \langle\nu_1|\bar{f}_{\lambda\mu;\alpha}(\mathbf{r})\,s_{\mu_2}|\nu\rangle \\
 &= -\sum_{\mu\mu_1\mu_2} (-)^\mu\,(1\,\mu_1\,1\,\mu_2|\lambda\,\mu)
 \sum_\alpha \sum_{\nu_1\,(\leq\varepsilon_\mathrm{F})}{\sum_\nu}^\sigma
 \langle\nu_1|\bar{f}_{\lambda\mu;\alpha}(\mathbf{r})\,s_{\mu_2}|\nu\rangle\,
 \langle\nu|f_{\lambda\mu;\alpha}(\mathbf{r})\,s_{\mu_1}|\nu_1\rangle \\
 &= -\sum_{\mu\mu_1\mu_2} (-)^{\lambda+\mu}\,(1\,\mu_2\,1\,\mu_1|\lambda\,\mu)
 \sum_\alpha \sum_{\nu_1\,(\leq\varepsilon_\mathrm{F})}{\sum_\nu}^\sigma
 \langle\nu_1|\bar{f}_{\lambda\mu;\alpha}(\mathbf{r})\,s_{\mu_2}s_{\mu_1}|\nu\rangle\,
 \langle\nu|f_{\lambda\mu;\alpha}(\mathbf{r})|\nu_1\rangle \\
 &= -\sum_\mu (-)^{\lambda+\mu} \sum_\alpha \sum_{\nu_1\,(\leq\varepsilon_\mathrm{F})}
 {\sum_\nu}^\sigma \langle\nu|f_{\lambda\mu;\alpha}(\mathbf{r})|\nu_1\rangle\,
 \langle\nu_1|\bar{f}_{\lambda\mu;\alpha}(\mathbf{r})\,
 [s\,s]^{(\lambda)}_\mu|\nu\rangle\,,
\end{split}\label{eq:eX3-sfr}\end{equation}
proving $\sum^\sigma_\nu\varepsilon^{(\mathrm{TN})}(\nu)=0$
on account of Eq.~(\ref{eq:ss}).


\end{document}